\documentstyle[aps,epsf]{revtex}
\begin{document}

\title{Color separate singlets in $e^+e^-$ annihilation}

\author{ 
Qun Wang$^{1}$
\thanks{on leave from Shandong University, China. 
email: qwang@public.sdu.edu.cn}, 
G\"osta Gustafson$^{1}$
\thanks{Gosta.Gustafson@thep.lu.se}, 
Qu-bing Xie$^{2}$
\thanks{xie@sdu.edu.cn}
}

\address{
$^1$ Department of Theoretical Physics, Lund University, 
S\"olvegatan 14A, S-22362 Lund, Sweden \\
$^2$ China Center of Advanced Science and Technology(World Lab)\\ 
P.O.Box 8730, Beijing 100080, P. R. China, and\\
Physics Department, Shandong University, 
Jinan, Shandong 250100, P. R. China\\
}

\maketitle

\begin{center}
{\large \bf Abstract}
\end{center}
We use the method of color effective Hamiltonian to study 
the properties of states in which a gluonic subsystem forms 
a color singlet, and we will study the possibility that such 
a subsystem hadronizes as a separate unit. A parton system 
can normally be subdivided into singlet subsystems in many 
different ways, and one problem arises from the fact that 
the corresponding states are not orthogonal. We show that
if only contributions of order $1/N_c^2$ are included, the 
problem is greatly simplified. Only a very limited number 
of states are possible, and we present an orthogonalization
procedure for these states. The result is simple and intuitive and
could give an estimate of the possibility to produce color 
separated gluonic subsystems, if no dynamical effects are 
important. We also study with a simple MC the possibility 
that configurations which correspond to "short strings"
are dynamically favored. The advantage of our approach over
more elaborate models is its simplicity, which makes it easier
to estimate color reconnection effects in reactions which
are more complicated than the relatively simple $e^+e^-$ annihilation. \\

%\begin{center}
%(\today )
%\end{center} 

%\newpage

\section{Introduction}
Hadronic processes in high energy reactions are generally 
formulated in terms of
two separate phases, one initial perturbative phase described by quarks
and gluons followed by a nonpertubative hadronization phase. In the
first phase the probability to produce a specific parton state, with
specified colors, can be calculated by perturbative 
quantum chromodynamics(PQCD). The transition from 
a partonic state to a hadronic one 
is a nonpertubative phenomenon and can only be described 
by hadronization models. 
Hence there is an intermediate interface between the perturbative 
stable partonic state and the hadronic one. 
To describe this interface it is not enough to know the 
momenta and colors of all partons, we also have to know how 
they are connected. Each color charge must be associated with a 
partner anticharge. Therefore the interface is normally described by a string or a  
cluster chain. This intermediate interface is needed not only to 
determine the hadronization, it is also needed to specify the 
termination of the perturbative cascade. 
In the HERWIG\cite{web,web1} and ARIADNE\cite{ariadne} 
cascades the termination is specified by a local transverse momentum 
and in the PYTHIA/JETSET\cite{sjopyth} 
cascade by the virtual mass of the parent. In all 
cases an ordering has to be known. We want to stress that in a 
perturbative calculation, which includes interference effects, the 
necessary parent-daughter relation cannot be specified from 
perturbative QCD. Instead it depends on the nonperturbative, soft, 
properties of the strong interaction. This implies that also if 
one assumes e.g. local parton hadron duality(LPHD), where the number 
of hadrons is directly related to the number of partons, a 
calculation of the multiplicity must rely on assumptions about the 
cutoff which cannot be specified from perturbative QCD alone.
(These features have been more thoroughly discussed in Ref.\cite{eden}.)

This problem does not appear in the large $N_c$ limit. When $N_c
\rightarrow \infty $ there is a single color ordering, which is the only
possible one. Planar Feynman diagrams dominate and many interference
terms vanish. In this limit we 
also find from perturbative QCD that generally partons are combined so
as to produce a rather "short" string or cluster chain. The process 
$e^+e^-\rightarrow q\overline{q}+ng$ corresponds to a specific
chain stretched by the gluons. In the perturbative phase, 
the only possibility to split the system
into two or more separate subsystems, is by the process $g\rightarrow
q\overline{q}$. Thus all strings or chains have a quark or an antiquark
in the ends, and it is not possible to isolate a purely gluonic
subsystem.

For finite $N_c$, there can be many different ways to connect the
color charges. This ambiguity is connected with the possibility for two
gluons to have identical colors, and is therefore generally of order 
$1/N_c^2$. 
New types of states are possible compared to the large $N_c$
limit. Thus for the $e^+e^-$ annihilation process it is possible to have a
purely gluonic subsystem, which can form a color singlet and therefore
hadronize as separated from the remainder. (This possibility was to our
knowledge first discussed in Ref.\cite{randa}.)
%This should be ref 7a

We are then facing the question: what decides the hadronization in a
situation when several different color connections are possible:
\begin{itemize}
\item Does Nature make a random choice among the different possibilities?
This means that the problem is determined purely by the SU(3) group
structure.
\item Are some states dynamically favoured? If this is the case the
probability for a specific configuration should depend on the
combination of a group factor and a dynamic factor.
\item If dynamics is important, are the properties in momentum space more
essential, (e. g. favouring states which correspond to "short strings"
Ref.\cite{friberg}) or are the properties in coordinate 
space more important (as proposed e. g. in Ref.\cite{torb,webb2})? 
\end{itemize}

We can isolate three different problems:
\begin{enumerate}
\item The SU(3) group structure - How many possibilities are there, and
what are their relative weights?
\item What are the dynamical properties of the alternative states allowed
by the group structure?
\item What observables can be used to solve the problems?
\end{enumerate}

In the standard models\cite{web,and,sjo} for parton 
cascades and hadronization, the only effect of 
finite $N_c$ originates from the coupling of a
gluon to quarks, which is given by $C_F = \frac{1}{2} N_c(1 - 1/N_c^2)$ and thus
reduced by a factor $(1 - 1/N_c^2)$ compared to the triple gluon coupling.
Only contributions from planar Feynman diagrams are included 
and a set of color suppressed effects are neglected. 
This implies that in these models, just as in the large $N_c$ limit, an
$e^+e^-\rightarrow q\overline{q}+ng$ event 
can split into two separate strings
or chains only via the process $g\rightarrow q\overline{q}$. Thus all
strings or chains have a quark or an antiquark in the ends, and it is
not possible to isolate a purely gluonic subsystem.

As mentioned above, the state intermediate 
between parton states and 
hadron states can be described by a string or cluster chain 
configuration. If this state contains a closed loop formed 
by gluons, we call it a color separate(CS) gluonic sub-singlet.

It is clear that to estimate the production rate of CS states, 
it is not enough to calculate the probability to find a 
gluonic subset which forms color singlet. For a state with 
many gluons this probability would always be one, and a 
gluonic singlet subsystem could also be formed in many 
different ways. 

Due to their non-perturbative nature, 
there is no way so far to determine the 
probability of string states with 
CS gluonic sub-singlets. 
A set of models have been presented, which include
nonconventional color configurations, 
where the partons can be "reconnected" compared to 
the "standard" configuration\cite{torb,webb2,leif,gus,si}. 
Such a model may easily appear as a
black box in the sense that it 
is difficult to judge what property of
the model is responsible for a 
particular observable effect. 
This paper is an attempt to make it 
easier to understand the features of 
CS states with gluonic sub-singlets. 
Here we use the method of 
the color effective Hamiltonian which is constructed 
from PQCD invariant amplitude\cite{wang} 
to study CS states corresponding to a particular 
string configuration 
from our available knowledge of PQCD. 

In section II- IV, we study the effects of 
the group structure using this method. 
We try to estimate the probability to 
produce a color singlet with gluonic sub-singlets(CSGS), 
disregarding any dynamic effects which might disfavor
its hadronization as a separate unit. 
For the parton states produced in PQCD 
the production amplitude can be written as 
a sum of terms, where each term contains a trace in 
color space, which corresponds to a definite color
ordering of the gluons. We call these parton states 
$|h_i\rangle$. We also want to study a different set of states, 
called singlet chain states, for which the color of a gluon (or quark)
and the anticolor of a neighbouring gluon (or antiquark)
form a color singlet. These states can form a single chain (the states
$|f_i\rangle$) or they can form two or more separate chains with at 
least one isolated gluonic loop (the states $|s_i\rangle$). 
In the large $N_c$ limit the states $|h_i\rangle$ are identical
to the states $|f_i\rangle$, and they correspond directly 
to string states with the same color ordering. For finite $N_c$
this is no longer the case. It is not possible to directly associate 
the singlet chain states, $|f_i\rangle$ and $|s_i\rangle$, to
definite string configurations. One reason is that the 
singlet chain states are not orthogonal. They can, however, 
as will be discussed in section IV, be adjusted into an orthogonal set by 
corrections of order $1/N_c^2$. We also show that to
leading order in $1/N_c^2$ the 
number of possible states $|s_i\rangle$ is significantly reduced,
which greatly simplifies the result.
The states obtained in this way 
may be associated with specified string (or chain) 
configurations. Here we propose a first estimate for the group 
weight corresponding to a specific string configuration obtained 
by projecting these string candidates upon a specific $|h_i\rangle$.

In section V we present some 
results from a very simple "reconnection" 
model based on the rules about leading CSGSs 
which we obtain in the earlier section. 
The comparison between global properties of events with 
and without CS states are given. 
By comparing with more elaborate models for $e^+e^-$ 
annihilation one may hope to clarify if some simple regularities are 
most important for the result. The simple model might then be used to
estimate reconnection effects in more complicated reactions.
 
\section{Color effective Hamiltonian}\label{h-c}

The QCD Lagrangian is:
$$
{\cal L}=-\frac 14F_{\mu \nu }^\alpha F^{\alpha \mu \nu }+\overline{q}
i\gamma ^\mu \partial _\mu q+g\overline{q}\gamma ^\mu A_\mu ^\alpha T^\alpha q 
$$
where $A_\mu ^\alpha $ $(\alpha =1,\cdots ,8)$ are eight gluon fields, $q$ is
the quark field with three colors $(R,Y,B)$; 
field tensor $F_{\mu \nu}^\alpha $ 
is given by
$$
F_{\mu \nu }^\alpha =\partial_\mu A_\nu ^\alpha -\partial_\nu A_\mu ^\alpha
+gf^{\alpha \beta \gamma }A_\mu ^\beta A_\nu ^\gamma 
$$
and $T^\alpha =\lambda ^\alpha /2$ where $\lambda ^\alpha $ is the Gell-Mann
matrix for $SU(3)$. We can write the QCD Lagrangian in another form by
combining Gell-Mann matrix to a set of new ones:
$$
\{\tau _{12}=\frac 12(\lambda _1+i\lambda _2),
\tau _{21}=\tau _{12}^{+},
\tau _{13}=\frac 12(\lambda _4+i\lambda _5),
\tau _{31}=\tau _{13}^{+},
\tau _{23}=\frac 12(\lambda _6+i\lambda _7),
\tau _{32}=\tau _{23}^{+},\lambda _3,\lambda _8\}
$$
Matrix $\tau _{ij}$ raises color $j$ to color $i$, and its Hermite conjugate 
$\tau _{ij}^{+}=\tau _{ji}$ has the opposite function. Accordingly we define
eight new gluon fields
$$
X_\mu ^3=\frac 1{\sqrt{2}}A_\mu ^3\;,\;X_\mu ^8=\frac 1{\sqrt{2}}A_\mu ^8 
$$
$$
X_\mu ^{21}=\frac 1{\sqrt{2}}(A_\mu ^1+iA_\mu ^2)\;,\;X_\mu ^{12}=\frac 1{
\sqrt{2}}(A_\mu ^1-iA_\mu ^2)\; 
$$
$$
X_\mu ^{31}=\frac 1{\sqrt{2}}(A_\mu ^4+iA_\mu ^5)\;,\;X_\mu ^{13}=\frac 1{
\sqrt{2}}(A_\mu ^4-iA_\mu ^5) 
$$
$$
X_\mu ^{32}=\frac 1{\sqrt{2}}(A_\mu ^6+iA_\mu ^7)\;,\;X_\mu ^{23}=\frac 1{
\sqrt{2}}(A_\mu ^6-iA_\mu ^7) 
$$
Hence the quark-gluon coupling part of the QCD Lagrangian can be written as: 
\begin{equation}
\label{l-q-g} 
\begin{array}{l}
{\cal L}_{int}=g\overline{q}\gamma ^\mu A_\mu ^\alpha T^\alpha q \\ =\frac g{
\sqrt{2}}\overline{q}\gamma ^\mu (\lambda ^3X_\mu ^3+\lambda ^8X_\mu ^8)q \\ 
+\frac g{\sqrt{2}}[\overline{q}\gamma ^\mu (\tau _{21}X_\mu ^{21}+\tau
_{31}X_\mu ^{31}+\tau _{32}X_\mu ^{32})q+h.c.] 
\end{array}
\end{equation}
where $X_\mu ^3$ couples to color $R\overline{R}$ and $Y\overline{Y}$ ; $
X_\mu ^8$ to $R\overline{R}$ ,$Y\overline{Y}$ and $B\overline{B}$; $X_\mu
^{21}$ to $R\overline{Y}$ ; $X_\mu ^{12}$ to $Y\overline{R}$; and so on. So
we write the QCD Lagrangian in a form with the quark-gluon interaction term
showing clear color significance. This triggers our following color treatment
for multigluon processes.

We can construct from PQCD a strict formulation to
calculate the cross section of color singlets of a multiparton system at the
tree level\cite{wang}. For the process 
$e^{+}e^{-}\rightarrow q\overline{q}+ng$, the
essential part of the formulation is to exploit the color effective
Hamiltonian $H_c$ to compute the amplitude $\left\langle f\right| H_c\left|
0\right\rangle $ of a certain color state $\left| f\right\rangle $. The
color effective Hamiltonian $H_c$ is found from the invariant amplitude 
$M_{ab}^{\alpha _1\alpha _2\cdots \alpha _n}$: 
\begin{equation}
\label{amp}M_{ab}^{\alpha _1\alpha _2\cdots \alpha _n}=\sum_P(T^{\alpha
_{P(1)}}T^{\alpha _{P(2)}}\cdots T^{\alpha _{P(n)}})_{ab}D^P 
\end{equation}
where $a/b$ is the color index for quark/antiquark; $\alpha _1,\alpha_2,
\cdots ,\alpha _n$ are those for $n$ gluons with $\alpha _u=1,\cdots ,8$
(for $u=1$,$\cdots $,$n$); the summation is over all permutations of 
$(1,2,\cdots ,n)$; $D^P\equiv D(q,\overline{q},g_{P(1)},g_{P(2)},
\cdots,g_{P(n)})$ is the function of parton momenta 
where momentum indices are
suppressed, and $P$ denotes a certain permutation of $(1,2,\cdots ,n)$; 
$(T^{\alpha _{P(1)}}T^{\alpha _{P(2)}}\cdots T^{\alpha _{P(n)}})_{ab}$ 
is the $a$-th row and the $b$-th column element of the matrix 
$(T^{\alpha_{P(1)}}T^{\alpha _{P(2)}}\cdots T^{\alpha _{P(n)}})$. 
The form of the invariant amplitude can be derived from 
the $SU(3)$ color structure of QCD.
A modern and convenient treatment was developed by Berends and Giele\cite
{berends}. They proposed a recursive method from which eq.(\ref{amp}) can be
readily obtained after including all Feynman diagrams at the tree level. 
Then $H_c$ is built from eq.(\ref{amp}) as follows: 
\begin{equation} 
\label{hc0} 
\begin{array}{c} 
H_c=\sum_P(T^{\alpha _{P(1)}}T^{\alpha _{P(2)}}\cdots T^{\alpha 
_{P(n)}})_{ab}D^P\Psi _a^{\dagger }\Psi ^{b\dagger }A_1^{\alpha _1\dagger
}A_2^{\alpha _2\dagger }\cdots A_n^{\alpha _n\dagger } \\ 
=\sum_P(1/\sqrt{2})^nTr(Q^{\dagger }G_{P(1)}^{\dagger }G_{P(2)}^{\dagger
}\cdots G_{P(n)}^{\dagger })D^P\qquad \qquad \quad 
\end{array}
\end{equation}
where the repetition of two subscripts represents summation(we use this
convention unless explicitly specified); $\Psi _a^{\dagger }=(R^{\dagger
},Y^{\dagger },B^{\dagger })$ is the color creation operator for quark; 
$\Psi ^{b\dagger }=(\overline{R}^{\dagger },\overline{Y}^{\dagger },\overline{
B}^{\dagger })$ is the anticolor creation operator for antiquark; 
$(Q^{\dagger })_a^b=\Psi ^{b\dagger }\Psi _a^{\dagger }$ is the nonet tensor
operator;  For gluon $u$, $G_u^{\dagger }$,
the gluon's color octet operator, is defined by: 
\begin{equation}
\label{g-t} 
G_u^{\dagger }=(1/ 
\sqrt{2})\lambda ^{\alpha _u}A_u^{\alpha _u\dagger }
=\Psi _u^{b\dagger }\Psi _{ua}^{\dagger }-\frac 13E\delta (a,b)\Psi
_u^{b\dagger }\Psi _{ua}^{\dagger }\equiv G_u^{\prime \dagger }-\frac
13E\cdot 1_u\qquad \qquad \qquad \qquad \qquad \qquad \qquad \quad 
\end{equation}
where $G_u^{\prime \dagger }\equiv \Psi_u^{b\dagger }
\Psi _{ua}^{\dagger }$, $1_u\equiv \delta (a,b)\Psi_u^{b\dagger }
\Psi _{ua}^{\dagger }$; $\Psi _u^{b\dagger }$ and $\Psi_{ua}^{\dagger }$ 
are $(\overline{R}^{\dagger },\overline{Y}^{\dagger }, 
\overline{B}^{\dagger })_u$ and $(R^{\dagger },Y^{\dagger },B^{\dagger })_u$
respectively; $E$ is the unit $3\times 3$ matrix; $\delta (a,b)$ is Kronecker
symbol: $\delta (a,b)=
\{ \begin{array}{c}1,\;a=b \\ 
0,\;a\neq b 
\end{array}$; 
$A_u^{\alpha _u\dagger }$ is the $\alpha $$_u$-th gluon color operator 
whose meaning resembles that of $X_\mu $ field. 
Note that we use the same symbol $A$ to denote these color operators as the 
gluon field in the former paragraph and this does not cause ambiguity 
because we no longer use the gluon field in the rest of the paper.

The color effective Hamiltonian is another expression of the $S$ matrix, so 
it is not Hermitian. The validity of $H_c$ can be verified by the 
calculation of the matrix element for the process $e^{+}e^{-}\rightarrow q 
\overline{q}+ng$. For the color initial state $\left| 0\right\rangle $ and
the final state $\left| f\right\rangle =\left| \Psi _{a^{\prime }}\Psi
^{b^{\prime }}A_1^{\alpha _1^{\prime }}A_2^{\alpha _2^{\prime }}\cdots
A_n^{\alpha _n^{\prime }}\right\rangle $, summing over the color indices 
$a^{\prime }$/$b^{\prime }$ for the quark/antiquark and those $\alpha 
_1^{\prime },\alpha _2^{\prime },\cdots ,\alpha _n^{\prime }$ for $n$ gluons, 
and carrying out all phase space integrals, we obtain: 
\begin{equation}
\begin{array}{l}
\int d\Omega \sum_f\left| \left\langle f\right| H_c\left| 0\right\rangle \right|^2
=\int d\Omega \left\langle 0\right| H_c^{\dagger }H_c\left| 0\right\rangle \\
=\int d\Omega \,\,M_{ab}^{\alpha _1\alpha _2\cdots \alpha _n} 
\cdot (M_{ab}^{\alpha _1\alpha
_2\cdots \alpha _n})^{*} \\
=\sigma _{tree}(e^+e^-\rightarrow q\overline{q}+ng)
\end{array}
\end{equation}
where $\int d\Omega$ denotes phase space integrals including 
all kinematic factors.
We can see that the calculation of the total cross section via $H_c$ 
returns to the ordinary form.

A color state of the multiparton system $q\overline{q}+ng$ is composed of 
color charges of $q,\overline{q}$ and $n$ gluons. It belongs to the color 
space: 
\begin{equation}
\label{csp}\underline{3}_q\otimes \underline{3}_{\overline{q}}^{*}\otimes 
\underline{8}_1\otimes \underline{8}_2\otimes \cdots \otimes \underline{8}_n 
\end{equation}
or a little larger space: 
\begin{equation}
\label{csp1}\underline{3}_q\otimes \underline{3}_{\overline{q}}^{*}\otimes 
\underline{3}_1\otimes \underline{3}_1^{*}\otimes \underline{3}_2\otimes 
\underline{3}_2^{*}\otimes \cdots \otimes \underline{3}_n\otimes \underline{3}_n^{*} 
\end{equation}
The effects of unphysical ''singlet-gluon'' state brought by enlarging the 
color space $\underline{8}_n$ to $\underline{3}_n\otimes 
\underline{3}_n^{*}(=\underline{1}_n\oplus \underline{8}_n)$ 
in (\ref{csp1}) can be 
eliminated in calculating the matrix element of $H_c$, i.e. 
projecting the state to $H_c\left| 0\right\rangle $, because
$$
\left\langle 1_u\cdot the\,rest\,of\,\,the\,color\,state\right| H_c\left|
0\right\rangle =0 
$$
where $u$ denotes a gluon. In other words, $\langle f\mid H_c\mid 0\rangle
=0 $ if $\left| f\right\rangle $ contains the singlet composed of the color
and its anticolor of a single gluon. There are many ways of reducing the
color space (\ref{csp}) or (\ref{csp1}). 
If a complete and orthogonal singlet set 
is denoted by $\{\left| f_k\right\rangle $, $k=1,2,\cdots \}$, we have: 
\begin{equation}
\left| f_k\right\rangle \left\langle f_k\right| =1\
\,\,(completeness),\qquad \left\langle f_k\mid f_{k^{\prime }}\right\rangle
=\delta _{kk^{\prime }}\,\,\,\,(orthogonality) 
\end{equation}
and 
\begin{equation}
\begin{array}{l}
\sum_k\sigma_k 
=\int d\Omega \sum_k\left| \left\langle f_k\right| H_c\left| 0\right\rangle \right|^2
=\int d\Omega \left\langle 0\right| H_c^{\dagger }\left| f_k\right\rangle \left\langle
f_k\right| H_c\left| 0\right\rangle \\
=\int d\Omega \,\,M_{ab}^{\alpha _1\alpha _2\cdots \alpha_n}
\cdot (M_{ab}^{\alpha _1\alpha _2\cdots \alpha _n})^{*} 
=\sigma _{tree}(e^{+}e^{-}\rightarrow q\overline{q}+ng)
\end{array}
\end{equation}
This property implies that the sum of the cross sections over all color
singlets in a complete and orthogonal singlet 
set for the system ($q\overline{q}+ng$) 
is equal to the total cross section at the tree level. 

\section{Examples of complete and orthogonal singlet sets}\label{coss}

\subsection{$e^{+}e^{-}\rightarrow q\overline{q}g$}

For the process with only one gluon in the final state, i.e. 
$e^{+}e^{-}\rightarrow q\overline{q}g_1$, the color space is: 
\begin{equation}
\label{qqg-csp0}\underline{3}_q\otimes \underline{3}_{\overline{q}
}^{*}\otimes \underline{8}_1 
\end{equation}
The enlarged color space is: 
\begin{equation}
\label{qqg-csp1}\underline{3}_q\otimes \underline{3}_{\overline{q}
}^{*}\otimes \underline{3}_1\otimes \underline{3}_1^{*} 
\end{equation}

There is only one way of reducing the space (\ref{qqg-csp0}). It 
corresponds to a unique singlet of the system $q\overline{q}g$. This
singlet is $\left| Tr(QG_1)\right\rangle $, where $(Q)_a^b$ is the color
nonet tensor of the quark pair and can be written as the irreducible octet
tensor $(Q^{\prime })_a^b$ plus its trace $1_Q\equiv Tr(Q)$: 
$Q=Q^{\prime}+\frac 131_QE$. Thus the complete singlet set 
is composed of only one singlet: 
\begin{equation} 
\label{qqg-sgt1}\{\left| Tr(QG_1)\right\rangle \}=\{\left| 
Tr(Q^{\prime}G_1)\right\rangle \} 
\end{equation}
where the equality is due to the fact that the gluon octet tensor is 
traceless, i.e. $Tr(G_1)=0$.

There are many ways of reducing the space (\ref{qqg-csp1}). Here we choose 
the following one that can lead to the singlet chain state: 
\begin{equation} 
\label{qqg-rd}(\underline{3}_q\otimes \underline{3}_1^{*})\otimes ( 
\underline{3}_{\overline{q}}^{*}\otimes \underline{3}_1)=(\underline{1}
_{q1}\oplus \underline{8}_{q1})\otimes (\underline{1}_{1\overline{q}}\oplus 
\underline{8}_{1\overline{q}}) 
\end{equation}
Corresponding to this reduction is the complete and orthogonal singlet set 
as follows: 
\begin{equation}
\label{qqg-sgt2}\{\frac 13\left| 1_{q1}1_{1\overline{q}}\right\rangle ,\frac 
1{\sqrt{8}}\left| 8_{q1}\otimes 8_{1\overline{q}}\right\rangle \} 
\end{equation}
where $1_{uv}\equiv \delta (a_u,b_v)\Psi _{a_u}\Psi ^{b_v}$, and $u$ and $v$
denote the parton, $a_u/b_v$ is the color/anticolor index for parton $u/v$;
and
$$
\left| 8_{uv}\otimes 8_{st}\right\rangle \equiv Tr(G_{uv}^{\dagger 
}G_{st}^{\dagger })\left| 0\right\rangle =\left| 
Tr(G_{uv}G_{st})\right\rangle 
$$
where $u$, $v$, $s$, $t$ are parton labels and $(G_{uv}^{\dagger 
})_a^b\equiv \Psi _v^{b\dagger }\Psi _{ua}^{\dagger }-\frac 13E\delta 
(a,b)\Psi _v^{b\dagger }\Psi _{ua}^{\dagger }$. The first singlet of 
(\ref{qqg-sgt2}) is defined as the singlet chain 
state. Generally a singlet chain state starts from the quark end, 
connects $n$ gluons one by one in an order and 
ends at the antiquark end. Each piece of the chain state 
is a color singlet formed by the color of one parton and the 
anticolor of the next parton in this order. 
The second singlet of (\ref{qqg-sgt2}) is the orthogonal one of the 
singlet chain state and is obtained from the reduction 
$\underline{8}_{q1}\otimes \underline{8}_{1\overline{q}}$.

\subsection{$e^{+}e^{-}\rightarrow q\overline{q}gg$}

The color space is: 
\begin{equation}
\label{qqgg-csp0}
\underline{3}_q\otimes \underline{3}_{\overline{q}
}^{*}\otimes \underline{8}_1\otimes \underline{8}_2 
\end{equation}
The enlarged color space is: 
\begin{equation}
\label{qqgg-csp1}\underline{3}_q\otimes \underline{3}_{\overline{q}
}^{*}\otimes \underline{3}_1\otimes \underline{3}_1^{*}\otimes \underline{3}
_2\otimes \underline{3}_2^{*} 
\end{equation}

Let us look at one reduction way for space (\ref{qqgg-csp0}):
\begin{equation}
(\underline{3}_q\otimes \underline{3}_{\overline{q}
}^{*})\otimes (\underline{8}_1\otimes \underline{8}_2)
= (\underline{1}_{q\overline{q}}\oplus \underline{8}_{q\overline{q}})
\otimes (\underline{8}_1\otimes \underline{8}_2)
\end{equation}
The complete and orthogonal singlet set 
corresponding to this reduction is:
\begin{equation} 
\label{cs1}
\{ (2\sqrt{6})^{-1}\left| 1_{q\overline{q}}(8_{1}\otimes 8_{2})\right\rangle ,
(3/80)^{1/2}\left| 8_{q\overline{q}}\otimes 
\{8_{1}\,,8_{2}\}\right\rangle ,(48)^{-1/2}\left| 8_{q\overline{q}}\otimes 
[8_{1}\,,8_{2}]\right\rangle \} 
\end{equation} 
where the symmetric octet state 
$\left| \{8_{1}\,,8_{2}\right\rangle\}$ 
and the anti-symmetric one $\left| [8_{1}\,,8_{2}]\right\rangle $ 
are defined as:
$$
\left| \{8_{1}\,,8_{2}\}\right\rangle =\left| \delta 
(a_1,b_2)(G_{1})_{a_1}^b(G_{2})_a^{b_2}+\delta 
(a_1,b_2)(G_{2})_{b_2}^b(G_{1})_a^{a_1}-\frac 23\delta 
(a,b)Tr(G_{1}G_{2})\right\rangle 
$$
$$
\left| [8_{1}\,,8_{2}]\right\rangle =\left| \delta 
(a_1,b_2)(G_{1})_{a_1}^b(G_{2})_a^{b_2}-\delta 
(a_2,b_1)(G_{2})_{a_2}^b(G_{1})_a^{b_1}\right\rangle 
$$
 
Another example is the reduction of the 
enlarged space (\ref{qqgg-csp1})
which can give the singlet chain state: 
\begin{equation}
\label{gg-rd} 
\begin{array}{l}
( 
\underline{3}_q\otimes \underline{3}_1^{*})\otimes (\underline{3}_1\otimes 
\underline{3}_2^{*})\otimes (\underline{3}_2\otimes \underline{3}_{\overline{
q}}^{*}) \\ =(\underline{1}_{q1}\oplus \underline{8}_{q1})\otimes ( 
\underline{1}_{12}\oplus 8_{12})\otimes (\underline{1}_{2\overline{q}}\oplus 
\underline{8}_{2\overline{q}}) 
\end{array}
\end{equation}
The complete and orthogonal singlet set corresponding to this reduction is: 
\begin{equation}
\label{gg-st} 
\begin{array}{l}
\{(3 
\sqrt{3})^{-1}\left| 1_{q1}1_{12}1_{2\overline{q}}\right\rangle ,(2\sqrt{6}
)^{-1}\left| 1_{q1}(8_{12}\otimes 8_{2\overline{q}})\right\rangle ,(2\sqrt{6}
)^{-1}\left| 1_{12}(8_{q1}\otimes 8_{2\overline{q}})\right\rangle , \\ (2 
\sqrt{6})^{-1}\left| 1_{2\overline{q}}(8_{q1}\otimes 8_{12})\right\rangle 
,(3/80)^{1/2}\left| 8_{2\overline{q}}\otimes 
\{8_{q1}\,,8_{12}\}\right\rangle ,(48)^{-1/2}\left| 8_{2\overline{q}}\otimes 
[8_{q1}\,,8_{12}]\right\rangle \} 
\end{array}
\end{equation}
In (\ref{gg-st}), the first singlet is the singlet chain state which has the 
parton configuration: $(qg_1g_2\overline{q})$. One can interchange gluon 
labels 1 and 2 in (\ref{gg-rd}) to obtain another singlet set which 
includes the configuration $(qg_2g_1\overline{q})$.

\section{CSGS in $q\overline{q}+ng$ system}
\label{cs-singlet}

As we see in the second section, the color effective Hamiltonian $H_c$ 
contains all perturbative dynamics relevant to color interaction. 
$H_c$ is actually the color $S$ matrix. 
We may define the $\left|in\right\rangle$ 
state as we do in quantum scattering theory: 
\begin{equation}
\left|in\right\rangle \equiv \left|H_c\right\rangle 
\equiv H_c\left|0\right\rangle
\end{equation}
We call the state $\left|H_c\right\rangle$ the physical color state. 
We can calculate the color matrix element of any color state 
$\left|f\right\rangle$ by projecting the state to 
the physical one: $\left\langle f\mid H_c\right\rangle$. 
According to (\ref{hc0}), the physical state is written as:
\begin{equation}   
\left|H_c\right\rangle =\sum_P(1/\sqrt{2})^n D^P 
\left| Tr(QG_{P(1)}G_{P(2)}\cdots G_{P(n)})\right\rangle 
\end{equation} 
where there are $n!$ trace terms. Each of them is associated with 
one order of gluons: $P(1,2,\cdots ,n-1,n)$, and it 
corresponds to a string connection of the same order at the large 
$N_c$ limit because they are orthogonal to each other. But with $N_c=3$, 
different trace terms overlap and no 
exact correspondence exists between 
a trace term with a specific gluon order in 
the physical color state at the partonic level 
and the string connection of the same order at 
the hadronic one. For each trace term, 
the color configuration is not uniquely determined 
and there are ingredients of color singlet chain states 
and CSGSs. In this section, we will discuss 
some properties of the CSGS and show how to 
estimate their probability from PQCD.

As the first example, let us look at the simplest case: 
$e^+e^-\rightarrow q\overline{q}g_1g_2$.
The physical color state for this process is:
\begin{equation}
\label{hc2g}
\left|H_c\right\rangle =\frac 12
\left| Tr(QG_1G_2)\right\rangle D^{12}+
\frac 12
\left| Tr(QG_2G_1)\right\rangle D^{21}
\end{equation} 
According (\ref{cs1}), the only CSGS for 
the system $q\overline{q}g_1g_2$ is:
\begin{equation}
\begin{array}{l}
\left|s_1\right\rangle =\frac 1{\sqrt{N_c}}\frac 1{\sqrt{N_c^2-1}}
\left| Tr(G_1G_2)1_{q\overline{q}}\right\rangle \\
=\frac 1{\sqrt{N_c}}\frac 1{\sqrt{N_c^2-1}}
(\left|1_{12}1_{21}1_{q\overline{q}}\right\rangle -
\frac 1{N_c}\left|1_{1}1_{2}1_{q\overline{q}}\right\rangle )
\end{array}
\end{equation}
where $\frac 1{\sqrt{N_c}}\frac 1{\sqrt{N_c^2-1}}$ is the normalization 
factor. The color part of the first trace term in the physical state is 
denoted by $\left|H_1\right\rangle$:
\begin{equation}
\begin{array}{l}
\left|H_1\right\rangle =\left| Tr(QG_1G_2)\right\rangle \\
=\left|1_{q1}1_{12}1_{2\overline{q}}\right\rangle -
\frac 1{N_c}\left|1_{1}1_{q2}1_{2\overline{q}}\right\rangle -
\frac 1{N_c}\left|1_{2}1_{q1}1_{1\overline{q}}\right\rangle +
\frac 1{N_c^2}\left|1_{1}1_{2}1_{q\overline{q}}\right\rangle 
\end{array}
\end{equation}
where $\left|H_1\right\rangle $ has the order $(g_1g_2)$. 
We can make the state 
$\left|H_1\right\rangle $ normalized:
\begin{equation} 
\left|h_1\right\rangle \equiv 
\frac{1}{\sqrt{N_c^3-2N_c+1/N_c}}\left|H_1\right\rangle 
\end{equation}
From now on, for the sake of convenience, 
we will simply call the color part of a trace term in the full physical 
state (\ref{hc2g}) a trace state. 
So the projection of the state $\left|s_1\right\rangle $ on 
the normalized trace state $\left|h_1\right\rangle $ is:
\begin{equation}
\left\langle s_1 \mid h_1\right\rangle \simeq \frac{1}{N_c}
\end{equation}
We should compare the above result with that of two 
singlet chain states 
$\left|f_1\right\rangle =\frac 1{\sqrt{N_c^3}}
\left|1_{q1}1_{12}1_{2\overline{q}}\right\rangle $ and
$\left|f_2\right\rangle =\frac 1{\sqrt{N_c^3}}
\left|1_{q2}1_{21}1_{1\overline{q}}\right\rangle $
which has the order $(qg_1g_2\overline{q})$ and 
$(qg_2g_1\overline{q})$ respectively. Their projections on 
the normalized trace state $\left|h_1\right\rangle $ 
are given by:
\begin{equation}
\begin{array}{l}
\left\langle f_1 \mid h_1\right\rangle =1\\
\left\langle f_2 \mid h_1\right\rangle =
\frac{1}{N_c^2}
\end{array}
\end{equation}
Hence we conclude that the CSGS accounts for 
about $1/N_c^2$ of the trace state $\left|h_1\right\rangle $ 
and there is a small portion ($1/N_c^4$) of 
$\left|f_2\right\rangle $ in $\left|h_1\right\rangle $ due to 
the different gluon order between the two states.

The projection of $\left|s_1\right\rangle $ on another trace 
term $\left|H_2\right\rangle =\left| Tr(QG_2G_1)\right\rangle $ 
is the same as on $\left|H_1\right\rangle $:
\begin{equation}
\left\langle s_1 \mid H_1\right\rangle =
\left\langle s_1 \mid H_2\right\rangle 
\end{equation}
since there is no order for two gluons in 
$\left| H_1\right\rangle$ and $\left| H_2\right\rangle$ :
$\left| Tr(G_1G_2)\right\rangle =
\left| Tr(G_2G_1)\right\rangle $.
So the projection of $\left|s_1\right\rangle $ on 
the full physical state $\left|H_c\right\rangle $ 
is given by:
\begin{equation}
\left\langle s_1 \mid H_c\right\rangle =
\frac{\sqrt{N_c-1/N_c}}{2}(D^{12}+D^{21})
\end{equation}
The probability for the parton system $q\overline{q}g_1g_2$ 
to be in $| s_1\rangle $ is obtained by:
\begin{equation}
\begin{array}{l}
Prob(\left|s_1\right\rangle )=
\frac{1}{\sigma _0}
\int d\Omega \left|\left\langle s_1 \mid H_c\right\rangle \right|^2 \\
=\frac{1}{\sigma _0}
\frac{N_c^2-1}{4N_c}\int d\Omega 
(\left|D^{12}\right|^2+\left|D^{21}\right|^2+
2Re(D^{12}D^{21*}))
\end{array}
\end{equation}
where $\sigma _0\equiv \sigma _{tree}
(e^+e^-\rightarrow q\overline{q}g_1g_2)$ is the tree level cross section 
for $e^+e^-\rightarrow q\overline{q}g_1g_2$.

There is only one CSGS for 
$q\overline{q}g_1g_2$. As the number of gluons increases, more 
such states arise. For the second example, we investigate the process:
$e^+e^-\rightarrow q\overline{q}g_1g_2g_3$. 
The full physical color state for the process is:
\begin{equation}
\begin{array}{l}
\left|H_c\right\rangle =\frac 1{(\sqrt{2})^3}[
D^{123}\left| Tr(QG_1G_2G_3)\right\rangle 
+D^{132}\left| Tr(QG_1G_3G_2)\right\rangle 
+D^{213}\left| Tr(QG_2G_1G_3)\right\rangle \\
+D^{231}\left| Tr(QG_2G_3G_1)\right\rangle 
+D^{312}\left| Tr(QG_3G_1G_2)\right\rangle 
+D^{321}\left| Tr(QG_3G_2G_1)\right\rangle ]
\end{array}
\end{equation}
Without loss of generality, we choose one trace term, e.g. 
$\left| H_1\right\rangle \equiv \left| Tr(QG_1G_2G_3)\right\rangle $ 
to go through our discussion. The other situations can be instantly 
obtained by permuting the gluon labels. 

As in the case of $e^+e^-\rightarrow q\overline{q}g_1g_2 $ , when 
$N_c$ is large, the normalized state 
$\left| h_1\right\rangle $ 
is approximately the same as that of the singlet chain state 
$\left|f_1\right\rangle \equiv 
(1/N_c^2)\left| 1_{q1}1_{12}1_{23}1_{3\overline{q}}\right\rangle $, 
noting that $\left| f_1\right\rangle $ 
has the same order of gluons as $\left| h_1\right\rangle $. 
There are five CSGSs for the system
$q\overline{q}g_1g_2g_3$:
\begin{equation}
\label{cs12345}
\begin{array}{l}
\{\left| s_i\right\rangle \;,\;i=1,2,3,4,5\}\\
=\{A\left| Tr(G_1G_2)1_{q3}1_{3\overline{q}}\right\rangle ,
A\left| Tr(G_2G_3)1_{q1}1_{1\overline{q}}\right\rangle ,
A\left| Tr(G_1G_3)1_{q2}1_{2\overline{q}}\right\rangle ,\\
B\left| Tr(G_1G_2G_3)1_{q\overline{q}}\right\rangle ,
B\left| Tr(G_1G_3G_2)1_{q\overline{q}}\right\rangle \}
\end{array}
\end{equation}
where A and B are normalization factors and given by:
\begin{equation}
A=\frac{1}{N_c\sqrt{N_c^2-1}}\;\;,\;\;B=\frac{1}{\sqrt{N_c^4-3N_c^2+2}}
\end{equation}
These singlets are all made up of two sub-singlets. 
For the first three states(i=1,2,3), one sub-singlet is 
formed by two gluons, the other is by the remaining gluon 
with the quark-antiquark pair, while for 
the last two states(i=4,5), all three gluons form one 
sub-singlet and the quark-antiquark pair forms the other one. 
Their projections on the normalized trace state 
$\left| h_1\right\rangle $ are:
\begin{equation}
\begin{array}{l}
\left|\left\langle s_1 \mid h_1\right\rangle\right|^2 
=\left|\left\langle s_2 \mid h_1\right\rangle\right|^2 \simeq 1/N_c^{2} \\
\left|\left\langle s_3 \mid h_1\right\rangle\right|^2 \simeq 1/N_c^6 \\
\left|\left\langle s_4 \mid h_1\right\rangle\right|^2 \simeq 1/N_c^2 \\
\left|\left\langle s_5 \mid h_1\right\rangle\right|^2 \simeq 1/N_c^6
\end{array}
\end{equation}
We see that the projection squares of 
$\left|s_1\right\rangle$, $\left|s_2\right\rangle$ and 
$\left|s_4\right\rangle$ 
on $\left| h_1\right\rangle $ are all about $1/N_c^2$ to the 
leading order in $N_c$, while those of $\left|s_3\right\rangle$ and 
$\left|s_5\right\rangle$ are suppressed by an additional factor $1/N_c^4$. 
Recalling (\ref{cs12345}), we may write down the parton orders 
for all these CSGSs below:
\begin{equation} 
\begin{array}{l}
(12)(03)\rightarrow \left|s_1\right\rangle \\
(01)(23)\rightarrow \left|s_2\right\rangle \\
(02)(13)\rightarrow \left|s_3\right\rangle \\
(0)(123)\rightarrow \left|s_4\right\rangle \\
(0)(132)\rightarrow \left|s_5\right\rangle \\
\end{array}
\end{equation}
where a parenthesis denotes a sub-singlet; 0 stands for 
the quark-antiquark pair and 1,2,3 for gluons. We 
compare these orders with the order in $\left|H_1\right\rangle$:
(0123). For $\left|s_1\right\rangle$, the orders in both sub-singlets 
are the same as (0123), while for, e.g. $\left|s_3\right\rangle$, 
the order in both sub-singlets are different from (0123)
because both 0 and 1 are not adjacent to 2 and 3 in (0123)
respectively. We conclude that the parton orders in 
$\left|s_1\right\rangle$, $\left|s_2\right\rangle$ and 
$\left|s_4\right\rangle$ have the least differences from (0123), 
while those in $\left|s_3\right\rangle$ and $\left|s_5\right\rangle$
are more different from (0123) than the former three states. 
In other words, each sub-singlet of $\left|s_1\right\rangle$, 
$\left|s_2\right\rangle$ and $\left|s_4\right\rangle$ 
conserves the parton order in the trace state at most, so they 
have the largest overlap with 
the trace state. The reason that 
$\left|s_3\right\rangle$ and $\left|s_5\right\rangle$ 
are more suppressed is that their sub-singlets do not all 
keep the parton order in the trace state. The inner product of 
any two states in $\left|s_1\right\rangle$, $\left|s_2\right\rangle$ and 
$\left|s_4\right\rangle$ is about $1/N_c^2$, while that 
of one state in $\left|s_1\right\rangle$, $\left|s_2\right\rangle$ and 
$\left|s_4\right\rangle$ with one of 
$\left|s_3\right\rangle$ and $\left|s_5\right\rangle$
is about $1/N_c^3$ or higher. 

This rule can be further verified for 
$e^+e^-\rightarrow q\overline{q}g_1g_2g_3g_4$. CSGSs 
with the leading contribution to 
$\left|H_1\right\rangle =\left|Tr(QG_1G_2G_3G_4)\right\rangle$ are:
\begin{equation}
\label{order}
\{(123)(04),(01)(234),(12)(034),(23)(014),(012)(34),(0)(1234)\}
\end{equation} 
Their projection squares on the normalized trace state 
$\left|h_1\right\rangle $ are 
all $1/N_c^2$ to the leading order in $1/N_c$. The inner product of 
any two different states in the above list is about $1/N_c^2$.
Note that we only include states with two sub-singlets 
because the projection squares on $\left|h_1\right\rangle $ 
for those with three or more sub-singlets 
are at least suppressed by additional $1/N_c^2$ as compared to 
those with only two. 

For cases of $n$-gluons where $n\leq 6$, 
the above conclusions can be verified by 
direct calculation. For cases of $n>6$, a formal proof 
is given below. Suppose any two different 
leading CSGSs have an inner product 
of $1/N_c^2$ to the leading order in 
$1/N_c$ in $n$-gluons case. Let us prove that the 
same statement holds for $(n+1)$-gluons case. 
We focus on two different CSGSs which we call A and B:
\begin{equation}
\begin{array}{l}
A:(I)(II)=(0,1,2,\cdots ,k_1,k_1+l_1+1,\cdots ,n+1)
(k_1+1,k_1+2,\cdots ,k_1+l_1)\\
B:(III)(IV)=(0,1,2,\cdots ,k_2,k_2+l_2+1,\cdots ,n+1)
(k_2+1,k_2+2,\cdots ,k_2+l_2)
\end{array}
\end{equation} 
where 0 stands for $q\overline{q}$ and the parentheses for 
sub-singlets. Each CSGS has a gluonic sub-singlet with $l_1$ 
and $l_2$ gluons, respectively. For $n>6$, at least a common 
piece of $[a_1,a_2,a_3]$ in two CSGSs must exist. 
Here $a_1$, $a_2$ and $a_3$ denote three 
consecutive gluons in a sub-singlet of one CSGS. When we calculate 
the inner product of A and B, we get a factor of $N_c^2$ to the leading
order in $N_c$ from this common piece. 
Corresponding to A and B in $(n+1)$-gluons 
case, there exist two CSGSs A' and B' in $n$-gluons case where 
all parts of A'/B' are the same as A/B except that the common piece 
$[a_1,a_2,a_3]$ is changed to $[a_1,a_2]$ 
($a_3$ as the new added gluon to A and B) or $[a_2,a_3]$ 
($a_1$ as the new added gluon to A and B) or $[a_1,a_3]$ 
($a_2$ as the new added gluon to A and B). 
Note that any one of three cases 
are equivalent to our proof. The contribution from the common 
piece, say $[a_1,a_2]$, to the inner product of A' and B' 
is $N_c$ to the leading order in $1/N_c$, while that from the common
piece $[a_1,a_2,a_3]$ to the inner product of A and B
is about $N_c^2$. The normalization factor of about 
$N_c^{-(n+2)}$ for $(n+1)$-gluons case and 
$N_c^{-(n+1)}$ for $n$-gluons one must also be taken into 
account. Considering contributions from both the common piece and 
the normalization factor, we find that the inner product of A 
and B is the same as that of A' and B' to the leading order in $1/N_c$.
The same strategy can be also used to prove 
that the inner product between any leading CSGS and 
its corresponding trace state is about $1/N_c$.

Hence we summarize the following rules for 
the CSGS with the leading contribution to a trace state:
\begin{itemize}
\label{rule}
\item The state has only two sub-singlets to the leading order in $1/N_c$. 
Its projection square on the 
trace state is about $1/N_c^2$.
\item Each sub-singlet keeps its parton order as 
that in the trace state.
\item The inner product of any two different leading contribution states 
is about $1/N_c^2$, while that of a leading state with a 
non-leading one is about $1/N_c^3$ or higher. 
\end{itemize}

According to the above rules, we can determine 
for the process 
$e^+e^-\rightarrow q\overline{q}g_1g_2\cdots g_{n-1}g_{n}$ 
the total number of CSGSs with 
the leading contribution to a trace state. 
We suppose this trace state corresponds to the trace term 
$\left|Tr(QG_1G_2G_3\cdots G_{n-1}G_{n})\right\rangle$. 
Hence, for the configuration 
$(q\overline{q}g_1g_2\cdots g_{n-1}g_{n})$, there are 
$(n-1)$ states with one sub-singlet formed by 2 gluons and 
the other by the remaining gluons with the quark-antiquark pair. 
Obviously, there are $(n-k+1)$ states with one 
$k$-gluon sub-singlet and the other formed by the 
remaining partons. Note that all these sub-singlets keep 
their parton orders as that in the trace state. 
Altogether, the total number is:
\begin{equation}
\label{cs-number}
1+2+3+\cdots +n-1=\frac 12n(n-1)
\end{equation}
We hereafter denote this number by $m=\frac 12n(n-1)$. 
  
Now we start to discuss how to estimate the total 
probability of the CSGS. One problem we are facing is 
that CSGSs are not orthogonal to each other. One cannot 
obtain the total probability by summing all individual 
probability values for each state. The non-orthogonality 
would make the total probability derived in this way 
different from its correct value because the contribution from 
the overlap of any two states would be 
counted multiply. So for non-orthogonal states, 
the orthogonalization is needed.
Let us illustrate the procedure more explicitly. 
If we have a set of non-orthogonal states:
\begin{equation}
\{\left|f_i\right\rangle ,i=1,2,\cdots l\}
\end{equation}
suppose we have finished the orthogonalization 
and find a set of orthogonal states as follows:
\begin{equation}
\{\left|f'_i\right\rangle ,i=1,2,\cdots l\}
\end{equation}
The total probability for the set of states is:
\begin{equation} 
\begin{array}{l}
Prob(\sum \left|f'_i\right\rangle )=
\frac{1}{\sigma _0}
\sum _i\int d\Omega \left|\left\langle f'_i 
\mid H_c\right\rangle \right|^2 \\
=\frac{1}{\sigma _0}\frac{1}{2^n}\sum _i
\int d\Omega \left|\sum _PD^P\left\langle f'_i 
\mid Tr(QG_{P(1)}G_{P(2)}\cdots G_{P(n)}
\right\rangle \right|^2
\end{array}
\end{equation}
where $\sigma _0\equiv \sigma _{tree}
(e^+e^-\rightarrow q\overline{q}g_1g_2\cdots g_n)$ 
is the tree level cross section 
for $e^+e^-\rightarrow q\overline{q}g_1g_2\cdots g_n$.

As stated above, it is necessary to orthogonalize 
the non-orthogonal CSGSs to 
give their total probability. But actually 
this job proves to be extremely difficult.
The greatest obstacle lies in  that the 
number of CSGSs (which is also 
the dimension of the transformation matrix 
from original states to orthogonal ones)
is huge when the gluon number is large, and 
finding the transformation matrix with this huge 
dimension is almost impossible. 
All possible CSGSs must 
be taken into account, not only those 
with the leading contribution to a specific 
trace term of the full physical 
state $\left|H_c\right\rangle$. The reason 
for this is simple:  $\left|H_c\right\rangle$ 
includes all trace terms each of which corresponds 
to one order of gluons, hence a non-leading singlet 
for one trace term is not necessarily still the 
non-leading one for other trace terms with different 
gluon orders. Another big obstacle is that 
calculating the momentum function $D^P$ is a 
very complicated task, especially for 
multigluon processes. To some degree, these two difficulties 
make the orthogonalization of more theoretical value. 

The orthogonalization is, however, 
really simple and heuristic in some cases.
Let us look at such a case. Assume that 
the physical state is fixed at one trace term, e.g. 
$\left|H\right\rangle \equiv 
\left|Tr(QG_1G_2G_3\cdots G_{n-1}G_{n})\right\rangle$, we now try 
to estimate the probability of the CSGS in this case. 
Before we orthogonalize these singlets, we make some 
further approximations:
\begin{itemize}
\item We only take into account the leading contribution singlets as 
determined by the rule on page \pageref{rule}. 
All the inner products of two different leading singlets 
are supposed to be the same $1/N_c^2$. 
In fact, they are the same to the leading order in $1/N_c$. 
\item There is freedom to choose the transformation matrix
(the definition is given below). 
Considering the symmetry of 
inner products assumed in the previous item, 
we choose a simplest form for 
the transformation matrix: all the diagonal elements are assumed 
to be equal, and so are the non-diagonal elements.
\end{itemize} 

Keeping these approximations in mind, we will proceed with 
the problem. Now we have $m$ non-orthogonal states:
\begin{equation}
\label{csst}
\{\left|s_i\right\rangle ,i=1,2,\cdots ,m\}
\end{equation}
where the inner product of any two states is given by:
\begin{equation} 
\label{cij}
C_{ij}=C_{ji}=\langle s_i\mid s_j\rangle =\left\{ 
\begin{array}{l}
1,\,\,\,\,\,\,\,\,\,\,for\,\,\,i=j \\ 
1/N_c^2,\,\,\,\,for\,\,i\neq j 
\end{array} \right.
\end{equation}
Our goal is to find $m$ orthogonal states based on (\ref{csst}). 
These states are denoted by: 
$$
\{\left| s_i^{\prime }\right\rangle ,\,\,i=1,2,\cdots ,m\} 
$$
They are related to original states (\ref{csst}) by a linear 
transformation: 
\begin{equation} 
\label{f'}
\left| s_i^{\prime }\right\rangle =U_{ij}\left| s_j\right\rangle 
\end{equation} 
where the transformation matrix $U$ is a real $m\times m$ matrix. 
It is not the only transformation which transforms 
the original non-orthogonal states to orthogonal 
ones. Different transformation matrices correspond to different 
orthogonal sets. Two transformation matrices are connected 
through a unitary transformation, but the transformation matrix itself 
is not unitary because it associates orthogonal states with 
non-orthogonal ones. Any transformation matrix is 
equivalent to others in calculating the probability. 
The inner product of two new states is: 
\begin{equation}
\label{orth1}
\langle s_i^{\prime }\mid s_j^{\prime }\rangle =\delta
(i,j)=\sum_{k,l}U_{ik}U_{jl}\langle s_k\mid s_l\rangle
=\sum_{k,l}U_{ik}U_{jl}C_{kl} 
\end{equation}
As we state in the above, from the symmetric 
property of $C_{kl}$ in (\ref{cij}), we choose a 
symmetric matrix $U$. We search for a solution for 
which all diagonal elements have the same 
value and also all non-diagonal elements are equal to each other: 
\begin{equation}
\label{uii1} 
\begin{array}{l}
U_{ii}=a \\ 
U_{ij}=b \,,\,\,\,\; for\,\,i\neq j 
\end{array}
\end{equation}
where $a$ and $b$ are real numbers. Thus (\ref{orth1}) becomes:
\begin{equation}
\label{uik}
\begin{array}{l}
\sum_{k,l}U_{ik}U_{il}C_{kl}=1\\
\sum_{k,l}U_{ik}U_{jl}C_{kl}=0\,\,,\,\,\,\,\,\,\,for\,\,i\neq j 
\end{array}
\end{equation}
Substituting (\ref{uii1}) into (\ref{uik}), we obtain: 
\begin{equation}
\label{es1}
\begin{array}{l}
a^2+(m-1)b^2+\frac 2{N_c^2}(m-1)ab
+\frac 1{N_c^2}(m-1)(m-2)b^2=1 \\
2ab+(m-2)b^2+\frac 1{N_c^2}a^2+
\frac 2{N_c^2}(m-2)ab+\frac 1{N_c^2}(m-2)(m-3)b^2=0 
\end{array}
\end{equation} 
We find the exact solution for (\ref{es1}):
\begin{equation}
\label{sol}
a^2=\frac{N_c^2}{(m^2-3m+2+(m-1)N_c^2)y^2+2(m-1)y+N_c^2}\;,\;\; 
b=ya 
\end{equation} 
where $y$ is given by:
\begin{equation}
\label{y-sol}
y=\frac{-(m+N_c^2-2)\pm \sqrt{(N_c^2+1)m+N_c^4-2N_c^2-2}}
{(m-2)(m+N_c^2-3)}
\end{equation}
We note that this solution is finite when $m \rightarrow 2$,
and therefore regular for all positive values of m. We choose the
solution with a positive sign, as this corresponds to a unit matrix, $U=1$, 
in the limit $N_c \rightarrow \infty$.
The orthogonal states are:
\begin{equation}
\label{prime-state}
\left| s_i^{\prime }\right\rangle =a\left| s_i\right\rangle 
+b\sum _{j\neq i} \left| s_j\right\rangle
\end{equation}
The overlap between the trace state $|h\rangle$ and the states 
$\left| s_j\right\rangle$ is given by
\begin{equation}
\sum _i|\langle s_i^{\prime}|h\rangle |^2
=\frac{m}{N_c^2}(a+(m-1)b)^2
\end{equation}
When $N_c$ is large we find $y\simeq -\frac {1}{2N_c^2}$ 
and $a^2\simeq 1$. Thus to order $1/N_c^2$ the probability behaves 
as $\frac{m}{N_c^2}$ which increases monotonously with increasing
number of gluons. To estimate the probability we introduce the 
singlet chain state $\left| f\right\rangle$ into the list 
of states to be orthogonalised. Note that 
$\left|f\right\rangle$ has the same gluon order as the 
chosen trace state $\left| h\right\rangle$ to which the CSGS 
\{$\left|s_i\right\rangle $\} or 
\{$\left| s_i^{\prime }\right\rangle $\} correspond. We define the new 
orthogonal states as follows: 
\begin{equation}
\label{orth-state}
\begin{array}{l}
\left| s_i^{\prime \prime}\right\rangle 
=\alpha \left| s_i^{\prime}\right\rangle +\beta \left| f\right\rangle \\
\left| f^{\prime }\right\rangle =\gamma \left| f\right\rangle +
\delta \sum_k\left|s_k^{\prime}\right\rangle 
\end{array}
\end{equation}
When $N_c$ is large, keeping only the leading terms in $N_c$, 
(\ref{prime-state}) becomes
\begin{equation}
\left| s_i^{\prime }\right\rangle \simeq \left| s_i\right\rangle 
-\frac {1}{2N_c^2}\sum _{j\neq i} \left| s_j\right\rangle 
\end{equation}
which implies $\left\langle f|s_i^{\prime }\right\rangle \simeq 1/N_c$. 
$\alpha , \beta , \gamma , \delta $ are four
parameters to be determined from the following orthogonality:
\begin{equation}
\label{orth11}
\begin{array}{l}
\left\langle s_i^{\prime \prime}|s_i^{\prime \prime}\right\rangle =1  \\
\left\langle s_i^{\prime \prime}|s_j^{\prime \prime}\right\rangle =0 
\;\;for\;i\neq j 
\nonumber \\
\left\langle f^{\prime }|f^{\prime }\right\rangle =1 \\
\left\langle f^{\prime }|s_i^{\prime \prime}\right\rangle =0 
\end{array}
\end{equation}
From the first and second equation of (\ref{orth11}), we obtain 
$\alpha =1$ and $\beta =-2/N_c $. From the last two 
equations of (\ref{orth11}), we get to order $1/N_c^2$
\begin{equation}
\gamma ^2 \approx 1-\frac{3m}{N_c^2}; \;\; 
\delta \approx \frac{\gamma}{N_c}
\end{equation}
In the large $N_c$ limit, the squared projections 
of $|f^{\prime }\rangle$ and 
$|s_i^{\prime \prime}\rangle $ on the corresponding 
normalized trace state $| h\rangle $, i.e. 
the probabilities of $|f^{\prime }\rangle$ and 
$|s_i^{\prime \prime}\rangle $ in $| h\rangle $,  
are given by:
\begin{equation}
\begin{array}{l}
P_{f'} \approx \left |\left\langle f^{\prime }|h\right\rangle \right |^2
= \gamma ^2(1+m/N_c^2)^2\simeq \frac{1}{1+\frac{m}{N_c^2}}\\
P_{s''}=\left |\left\langle s_i^{\prime \prime }|h\right\rangle \right |^2
\approx 1/N_c^2 \simeq \frac{1}{N_c^2(1+m/N_c^2)}
\end{array}
\end{equation}
In the last aproximations we have added terms of higher order in $1/N_c^2$
in such a way that the total probability adds up to one for all values of $m$.
Thus the total probability for all $|s''\rangle $ states 
becomes $mP_{s''}\simeq \frac{m/N_c^2}{1+m/N_c^2}$, which increases 
monotonously with growing gluon number (or $m$). Correspondingly 
$P_{f'}$ decreases monotonously with increasing $m$. 

Summarily, what we have derived gives us 
some valuable messages about the properties of the CSGS. 
Here are some of them:
(a) The probability of color 
separate singlets cannot be neglected. They may 
account for a considerable fraction of the total events.
We try to estimate the probability to 
produce a CSGS, disregarding any 
dynamic effects which might disfavor
its hadronization as a separate unit. 
(b) The set of states corresponding 
to different string configurations must 
necessarily be orthogonal, as they 
correspond to different values for 
observables. In the large $N_c$ limit, CSGSs 
do not appear. A trace state is just the singlet chain state 
having the same gluon order. All singlet chain states
are orthogonal to CSGSs. 
For finite $N_c$, neither are the three sets of states, 
trace states, singlet chain states and 
color separate states, orthogonal among themselves, nor 
is a set of states orthogonal to any other. 
These states are not possible candidates to 
represent pure string states. Keeping only 
the first correction term in an expansion in 
$1/N_c^2$, it is, however, possible to construct modified 
orthogonal states which corresponds to a specific 
trace state $|h_i\rangle$: 
$\{f_i^{\prime},\{s_{ij}^{\prime \prime}\}\}$.
Thus they form an orthogonal set of 
states, which may be associated with specified string (or chain) 
configurations. Here we propose a first estimate for the group 
weight corresponding to a specific string configuration obtained 
by projecting these string candidates upon a specific $|h_i\rangle$.
(c) For a trace state, 
leading CSGSs obey the rules on page \pageref{rule}.
In particular the rules are crucial 
for us to build a practical model 
to describe the production of color separate states. 
This is the subject which we will discuss in the next section.
 
\section{A color separate state model}

According to the arguments in the last section, 
the total percentage of color separate states should not 
be negligibly small, especially in multigluon 
process. It is of significance 
to introduce CS states in the current 
event generators to see how they influence the event shape and 
properties. In this section, we will build such a practical 
model based on and implementing the rules which 
we derived in the former section. 

Firstly, the difference between two terms, 
the CSGS and CS state, should be
clarified. We mention in the introduction that 
to describe the interface between the perturbative 
stable partonic state and the hadronic one, 
it is not enough to know the momenta and colors 
of all partons, we also have to know how 
they are connected. 
Their connection can be described by a string or cluster chain 
configuration where each color charge must be associated with a 
partner anticharge. 
The connection or the string/chain 
ordering cannot be specified by perturbative QCD alone. 
It depends on the nonperturbative, soft, 
properties of the strong interaction. 
If the string or cluster chain state contains a closed loop formed 
by gluons we call it a CS gluonic sub-singlet  
which is a non-perturbative object, while 
a CSGS is defined by the colors of partons 
which is of perturbative nature. A CSGS is not necessarily
a CS state. In this section, we
use JETSET to implement the fragmentation which
is a non-perturbative soft process. Hence we 
start using the term "CS state" 
in most places to describe the true objects formed at the
end of the parton cascade or at the beginning of the
hadronization. 

The outline of the model is as follows.
\begin{enumerate}

\item  We use a parameter $R$ to describe the total percentage of CS 
states in all events. We use JETSET to produce a configuration of partons 
where for the singlet or for each sub-singlet of the type 
$qg_1g_2\cdots g_l\overline{q}$, the color flow starts at the $q$, 
connecting gluons one by one and ends at $\overline{q}$. This 
color connection corresponds to a trace term in 
the physical color state $\left|H_c\right\rangle$ which we 
call a trace state.

\item We only consider those CSGSs with leading contributions 
to the trace state. For a configuration 
$qg_1g_2\cdots g_l\overline{q}$, there are $m=\frac{1}{2}l(l-1)$ 
leading states according to (\ref{cs-number}). 
Among them there are $(l-1)$ CSGSs of 2-gluon type, 
$(l-2)$ states of 3-gluon type,...,$(l-k+1)$ states of 
$k$-gluon type, etc.. Any two different CSGSs have 
approximately the same inner product($\simeq 1/N_c^2$). Each 
state's square of projection on the trace state(corresponding to 
the current color configuration) is the same 
($\simeq 1/N_c^2$). So all CSGSs play an equal role.  
However, not all of these states finally contribute to the 
event. Thanks to the fact that JETSET records the whole branching 
history for all partons, we can exclude those CSGSs 
where all gluons in the gluon-type sub-singlet are emitted from a single 
octet gluon. After filtered by this procedure, suppose there 
are $m'$ leading CSGSs remaining.

\item We consider two kinds of weights to select a CS 
state corresponding to the given configuration. 
One is the constant weight, i.e. 
each CS state is selected with equal probability. 
This is a natural choice if we only consider the color part and 
neglect the momentum or phase space sector.
The other is T(or $\lambda$) 
measure weight(we will come to this topic later in details). 
For each CS state, we have a T-weight. So there are $m'$ T-weights 
corresponding to $m'$ CS states. A CS state for 
the current configuration is selected according to these weights.

\item The fragmentation of the selected CS state is implemented by 
the Lund string model. The difference between 
an ordinary state of the type $qg_1g_2\cdots g_l\overline{q}$ and 
a CS state is that the former state forms an open string while the 
latter forms a closed string and an open one.

\end{enumerate}

In order to better understand our Monte Carlo simulation results, 
we should use a little more space for illustrating the physical meaning of 
the T-measure. We will address two subjects: (a) How is the T-measure 
defined and what's its physical meaning? (b) Its relationship 
to the $\lambda$-measure in the Dipole Cascade Model(DCM). 

\begin{figure}
\vspace*{1in}
\setlength{\epsfxsize}{6in}
\centerline{\epsffile{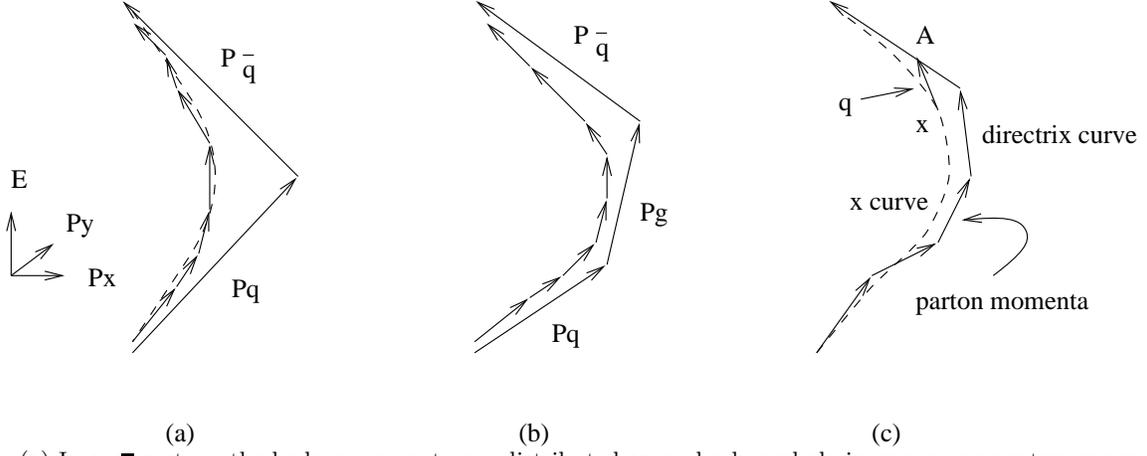}}
%\vspace*{.5in}
\caption{(a) In a $q\overline{q}$ system, the hadron momenta 
are distributed around a hyperbola in energy-momentum space.
(b) For a $q\overline{q}g$ system, the hadron momenta 
are distributed around two hyperbola. (c) For a multigluon state, 
the hadron momenta are distributed around a curve in 
energy-momentum space(called the x-curve) which smoothly 
follows the (color-ordered) parton momenta.}
\label{x-curve}
%\vspace*{.4in}
\end{figure}

For a $q\overline{q}$ system the produced hadrons are evenly distributed 
in rapidity, which means that their energy-momentum four-vectors, 
when plotted one by one, are distributed around a 
hyperbola as seen in Fig.\ref{x-curve}a. 
For a $q\overline{q}g$ system, the 
two string pieces will produce hadrons such that the momentum 
four vectors lie around two hyperbolae(Fig.\ref{x-curve}b). 
This corresponds to three jets along the parton momenta, 
and the jets are smoothly connected to each other. For a 
multigluon system it is possible to generalize the hyperbolae 
in the $q\overline{q}$ and  $q\overline{q}g$ cases and define 
a timelike curve in energy-momentum space. This curve 
(which we call the x-curve) follows the color ordered parton 
momenta in such a way that the corners are smoothed out with 
a resolution power given by a parameter $m_0$. 
The directrix curve $A_{\mu}$ for the string parton state 
is obtained by placing 
the four-momenta of all the partons one by one 
according to their color ordering, see Fig.\ref{x-curve}c. 
When $m_0\rightarrow 0$, the x-curve becomes the directrix.
Let $\xi$ be a parameter which runs along the directrix. To every 
point $A_{\mu}$ on the directrix there is a corresponding 
point $x_{\mu}$ on the x-curve. So the x-curve 
which is determined through 
the differential equations\cite{andersson}:
\begin{equation}
\begin{array}{l}
dx_{\mu}(\xi )=\frac{1}{m_0^2}(q\cdot dA)q_{\mu}\\
x_{\mu}(0)=A_{\mu}(0)=0
\end{array}
\end{equation}
where $q_{\mu}(\xi)\equiv A_{\mu}(\xi)-x_{\mu}(\xi)$ is 
tangent to the x-curve. T is defined by:
\begin{equation}
\begin{array}{l}
d(lnT)=\frac{q\cdot dA}{m_0^2}=\frac{\sqrt{dx^2}}{\sqrt{q^2}}\equiv d\lambda \\
T=exp(\lambda)
\end{array}
\end{equation}
Note that T is just the exponential of the area 
spanned by the x-curve and the directrix. 
T can also be interpreted as the exponential of 
the invariant length of the x-curve which we call the $\lambda$-measure.
The square amplitude to obtain a certain string 
state with $n$ partons is given by $\frac{1}{T_n(k_1,k_2,\cdots ,k_n)}$
where $k_1,k_2,\cdots ,k_n$ are four-momenta of $n$ partons. 
There is a recursive expression for $T_n$:
\begin{equation}
\label{qj1}
\begin{array}{l}
q_{j+1}=\gamma _{j+1}q_j+\frac 12(1+\gamma _{j+1})k_{j+1}\\
\gamma _{j+1}=1/[1+\frac{q_j\cdot k_{j+1}}{m_0^2}]\\
T_n^{-1}=\gamma _2\gamma _3\cdots \gamma _n
\end{array}
\end{equation}
The T-measure defined as above depends on the initial value of $q_{\mu}$. 
It is possible to start with the initial value $q_0=0$ for an 
open string. We regard the sequence 
$k_1,k_2,\cdots ,k_n$ as the first half of a full directrix period, 
the second half is $k_n,k_{n-1},\cdots ,k_1$, which is in the 
opposite order. This is one period: $k_1,k_2,\cdots ,k_n$,
$k_n,k_{n-1},\cdots ,k_1$.  After several periods of 
recursive calculation according to (\ref{qj1}), we 
will get for a period a stable $T$ which is insensitive to the 
initial value $q_0$. For a closed string where there is no obvious beginning, 
as a choice, we also use this strategy to get a stable $T$. 

In the string fragmentation model, 
the $\lambda$-measure is the effective rapidity 
range given by: $\lambda \sim \sum ln[(k_i+k_{i+1})^2/m_0^2]$
Hence, according to the weight defined by 
$\frac 1T=exp(-\lambda)$, we may tell which string parton state is 
"short" and which is "long". Here a "short" string means the string with 
small $\lambda$. The weight implies that 
a short string parton state is more favored than a long one. 

For a configuration $(qg_1g_2\cdots g_n\overline{q})$ produced 
by JETSET, suppose an allowed CS state is made up of 
a separate sub-singlet $(g_ig_{i+1}\cdots g_{i+j-1})$ of the 
j-gluon type and a sub-singlet 
$(qg_1\cdots g_{i-1}g_{i+j}g_{i+j+1}\cdots g_n\overline{q})$
of the ordinary type. The total weight is:
\begin{equation}
\begin{array}{l}
\frac{1}{T}=\frac{1}{T_1(k_i,k_{i+1},\cdots ,k_{i+j-1},k_i)
T_2(k_q,k_1,\cdots ,k_{i-1},k_{i+j},k_{i+j+1},\cdots ,k_n,k_{\overline{q}})}\\
\equiv exp(-\lambda _1-\lambda _2)
\end{array}
\end{equation}
There is a difference between $T_1$ and $T_2$:
$T_1$ is the T-measure for a closed string, so it is written as:
\begin{equation}
T_1=\gamma _{k_{i+1}}\gamma _{k_{i+2}}\cdots 
\gamma _{k_{i+j-1}}\gamma _{k_i}=exp(-\lambda _1)
\end{equation}
where the recursive sequence $(k_i,k_{i+1},\cdots ,k_{i+j-1},k_i)$ 
starts at $k_i$ to $k_{i+j-1}$ 
and then back to $k_i$; and $\lambda _1$ is approximately:
\begin{equation}
\lambda _1\sim ln(s_{i,i+1}/m_0^2)+ln(s_{i+1,i+2}/m_0^2)+\cdots 
+ln(s_{i+j-2,i+j-1}/m_0^2)+ln(s_{i+j-1,i}/m_0^2)
\end{equation}
where $s_{ij}\equiv (k_i+k_j)^2$ and the last term 
is the additional term compared to an open string. 
There is no such a term for an open string formed by the same sequence of 
momenta. We also use the $\lambda$-measure for the full 
configuration $(qg_1g_2\cdots g_n\overline{q})$ without 
interruption by separate sub-singlets: 
\begin{equation}
\lambda _0 \sim ln(s_{1,2}/m_0^2)+ln(s_{2,3}/m_0^2)+\cdots 
+ln(s_{n-2,n-1}/m_0^2)+ln(s_{n-1,n}/m_0^2)
\end{equation} 
We can compare $\lambda _0$ with $\lambda _1+\lambda _2$ by investigating 
their difference $(\lambda _1+\lambda _2)-\lambda _0$:
\begin{equation}
\label{l1l2l0}
\begin{array}{l}
\lambda _1+\lambda _2-\lambda _0 \\
\sim ln(s_{i+j-1,i})+ln(s_{i-1,i+j})-ln(s_{i-1,i})-ln(s_{i+j-1,i+j})\\
=ln(\frac{(k_{i+j-1}\cdot k_i)(k_{i-1}\cdot k_{i+j})}
{(k_{i-1}\cdot k_i)(k_{i+j-1}\cdot k_{i+j})})
\end{array}
\end{equation}
where $k_{i-1}$ and $k_i$, $k_{i+j-1}$ and $k_{i+j}$ are 
momenta of two gluons in neighborhood in the color 
flow respectively, while $k_{i+j-1}$ and 
$k_i$, $k_{i-1}$ and $k_{i+j}$ are 
not adjacent. Normally, we have:
\begin{equation}
(k_{i+j-1}\cdot k_i)(k_{i-1}\cdot k_{i+j})>
(k_{i-1}\cdot k_i)(k_{i+j-1}\cdot k_{i+j})
\end{equation}
which implies $\lambda _1+\lambda _2>\lambda _0$. That is to say:
the length of a CS state is usually longer than the ordinary state 
without interruption. We expect that the shorter the length of 
a string state is, the more favored it would be. We can find 
some configurations for a CS state 
where its length $\lambda _1+\lambda _2$ 
is shorter. It is conceivable that these 
configurations are more favored and have 
larger production weights. For example, 
if all the $j$ gluons 
in the separate sub-singlet plus two neighbor gluons to its two ends 
are soft, then the weight must be large because separating the $j$-gluon 
as a sub-singlet will not change the original $\lambda _0$ very much.

\begin{figure}
%\vspace*{1in}
\setlength{\epsfxsize}{4in}
\centerline{\epsffile{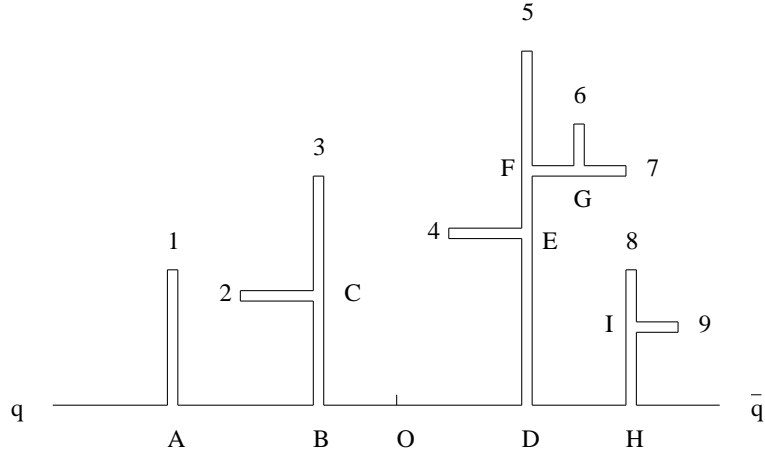}}
%\vspace*{.5in}
\caption{Diagram of the effective 
rapidity range (i.e. $\lambda$) 
for a parton configuration.}
\label{fold-line}
%\vspace*{.4in}
\end{figure}

\begin{figure}
%\vspace*{1in}
\setlength{\epsfxsize}{6in}
\centerline{\epsffile{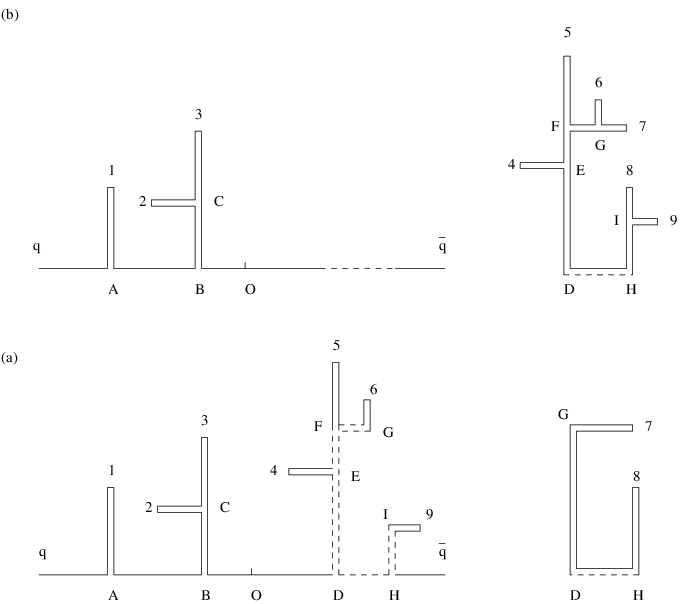}}
%\vspace*{.5in}
\caption{Diagram of the effective 
rapidity range (i.e. $\lambda$) 
for CS states. 
(a) $(q1234569\overline{q})(78)$
(b) $(q123\overline{q})(456789)$}
\label{cs-rapidity}
%\vspace*{.4in}
\end{figure}

To see the situation more easily and intuitively, 
we resort to the $\lambda$ 
diagram which is often used in the DCM. Fig.\ref{fold-line} 
is the diagram of the effective 
rapidity range (i.e. $\lambda$) for a 
parton configuration\cite{andersson1}. 
In this configuration, there are nine gluons. The order along the 
color flow is shown in the figure which reads 
$qg_1g_2\cdots g_9\overline{q}$. The length of the folded line 
$q123456789\overline{q}$ is $\lambda _0$. The length of the 
folded line between two neighboring gluons, e.g. $i$ and $i+1$, 
is $ln(s_{i,i+1}/m_0^2)$. In the figure, there are 
11 ends labeled by $q,\overline{q}$ and nine gluons respectively. 
We call these ends twigs. Along the line $q\overline{q}$, there are 
5 branches. In the branch which starts at D, there are 
4 twigs, two of which, 6 and 7, are from the sub-branch starting at F.
For this parton configuration, we will look at 
two examples of CS states to see 
how their $\lambda$ measures differ from $\lambda _0$. 
The first one is: $(q123\overline{q})(456789)$, which 
is shown in Fig.\ref{cs-rapidity}a. 
The length of the dashed line shows the $\lambda$ surplus compared 
to $\lambda _0$. The surplus is denoted by $\delta \lambda$: 
$\delta \lambda=2|DH|$. The second CS state is: 
$(q1234569\overline{q})(78)$ shown in 
Fig.\ref{cs-rapidity}b. The $\lambda$ surplus is: 
$\delta \lambda=2(|FG|+|DF|+|HI|+|DH|)$. 
Of course, the first CS state is more 
favored than the second one though it has more gluons in its 
gluon sub-singlet. If we look 
at them more carefully, we will find that the gluon sub-singlet 
of the first CS state is made up of all gluons from two adjacent 
branches, while the sub-singlet of the second is formed 
by part of gluons in the two branches. We also see that 
the shorter the line $|DH|$ is, the larger production weights 
the CS states have. Here $|DH|$ is just the rapidity 
difference between the two adjacent branches. We see that 
those CS states with their gluon sub-singlets formed by adjacent 
twigs (note that twig is the smallest branch) have large 
production rate. In this case most gluons in gluon sub-singlets 
are soft ones. 

We summarize the above arguments 
that most of CS states 
produced in the process have the following configurations: 
\begin{itemize}
\item The gluon sub-singlet of 
a CS state is made up of a series of neighboring gluons. 
If part of a branch joins the sub-singlet,  
the rest of this branch should also join it. 
Any separation of one part from the rest of the same branch would 
result in the suppression of the production weight. The whole 
branch is like a skeleton. The principle says, if the skeleton is in, 
the whole body should be in. 
\item The rapidity difference between 
two branches at the two ends of the continuous branch sequence 
is as small as possible. This implies 
that all branches are as far from the root 
(or beginning) of the cascade tree as possible. Hence 
those CS states with their gluon sub-singlets formed by adjacent 
twigs have large production rate.
\end{itemize}

\begin{figure}
%\vspace*{1in}
\setlength{\epsfxsize}{6in}
\centerline{\epsffile{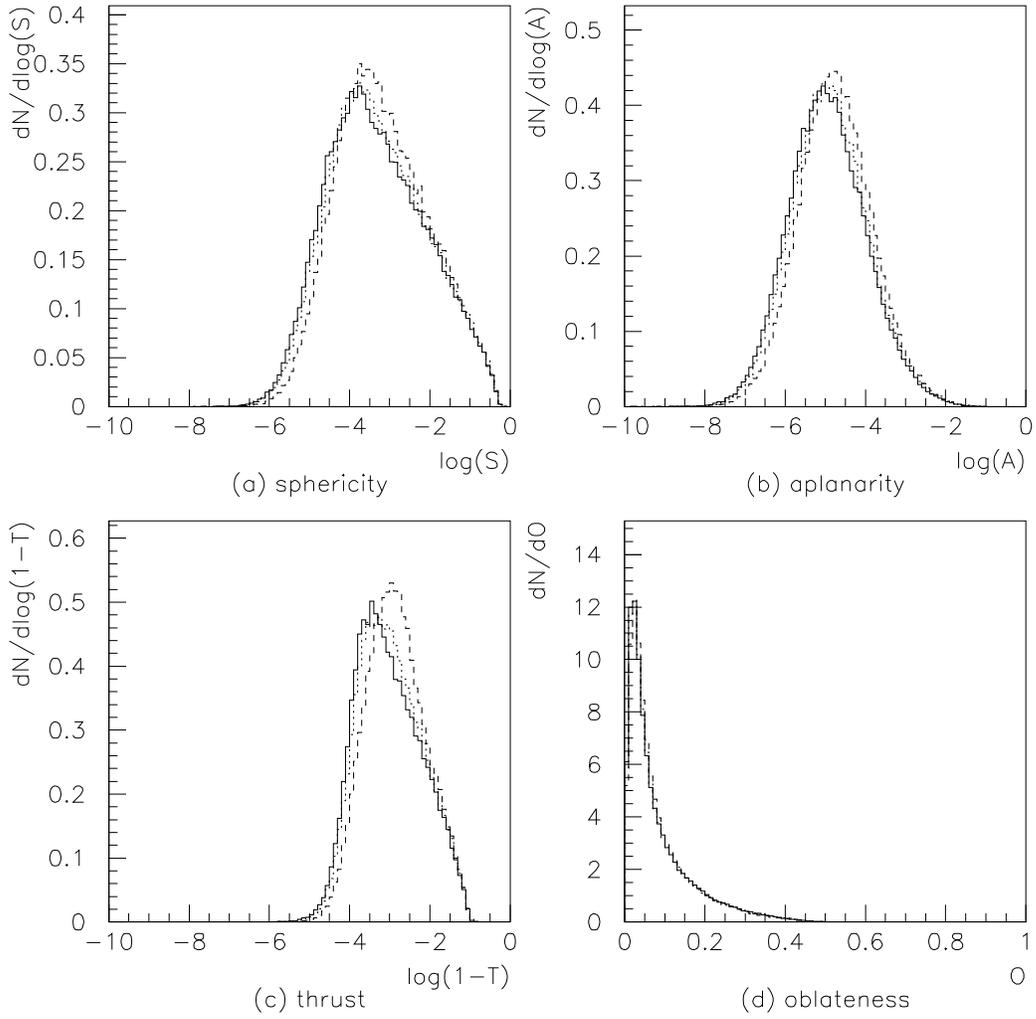}}
%\vspace*{.5in}
\caption{This figure is for comparing global quantities of events 
(sphericity, aplanarity, thrust and oblateness) 
among cases where CS states are produced with 0\%, 100\% and 30\% probability. 
The CS states are chosen with constant (or equal) weight. 
The energy is 91 GeV. Solid line: no CS states; 
Dashed line: 100\% CS states. Dotted line: 30\% CS states. }
\label{fig:1}
%\vspace*{.4in}
\end{figure}

\begin{figure}
%\vspace*{1in}
\setlength{\epsfxsize}{6in}
\centerline{\epsffile{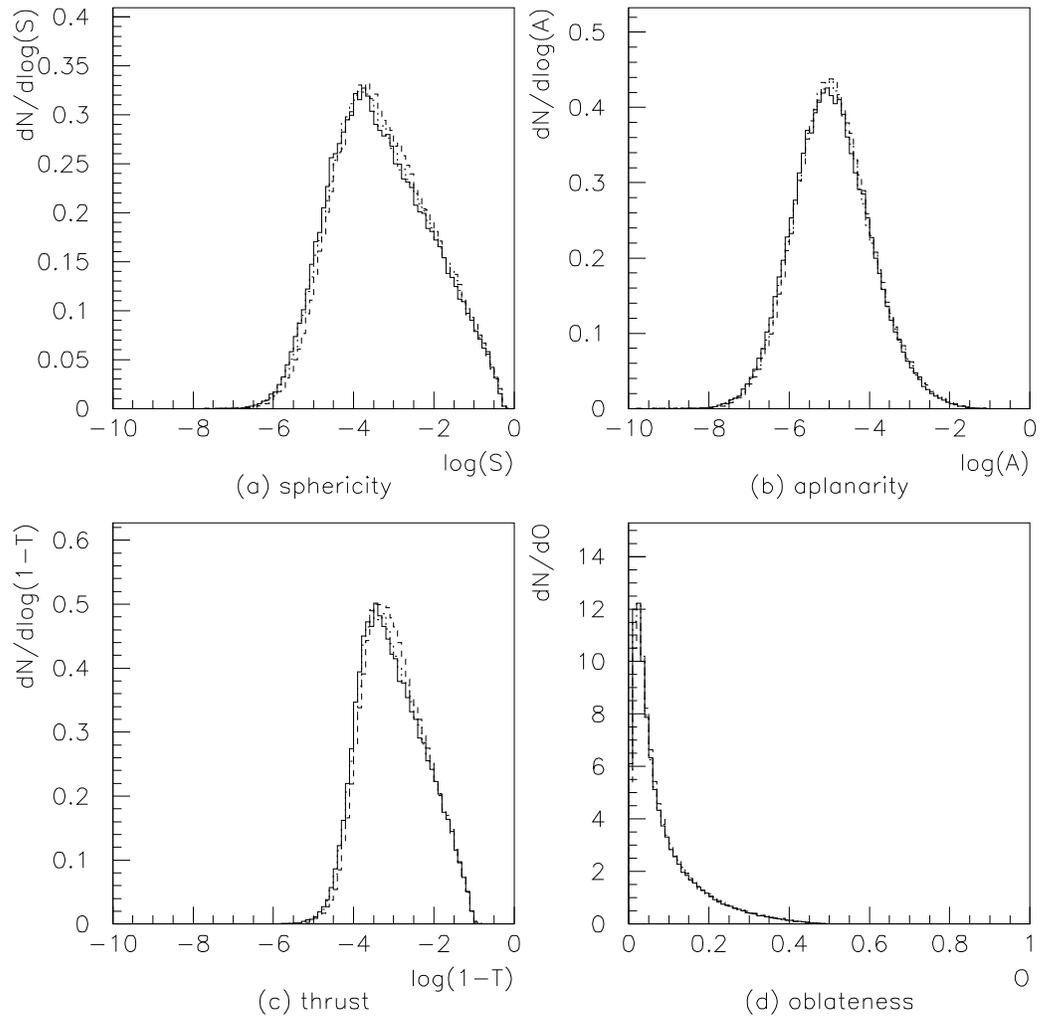}}
%\vspace*{.5in}
\caption{Same as Fig.\ref{fig:1} except that the CS states are 
chosen with T-weight. }
\label{fig:2}
%\vspace*{.4in}
\end{figure}

\begin{figure}
%\vspace*{1in}
 \setlength{\epsfxsize}{6in}
  \centerline{\epsffile{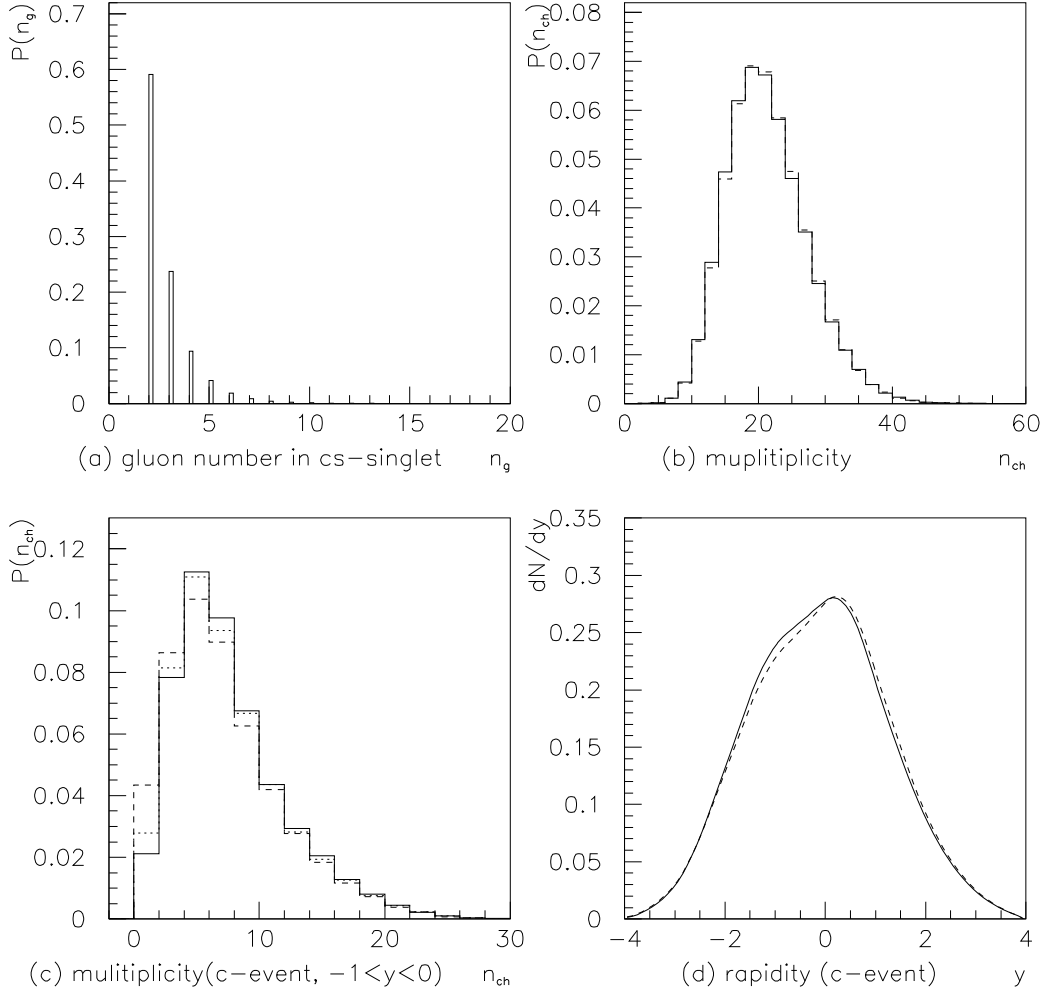}}
%\vspace*{.5in}
\caption{
The energy is 91 GeV. T-weight is used to select CS states. 
(a) The probability distribution of 
gluon number in the gluon type sub-singlet of the CS state. 
(b) Multiplicity distribution of charged particles. 
Solid line: no CS states; Dashed line: 30\% CS states. 
(c) Multiplicity distribution of charged particles 
in the rapidity range $-1<y<0$ for c-events 
(i.e. the initial quark-aniquark pair produced at the electroweak 
vertex is $c\overline{c}$). 
The rapidity axis is chosen along the added $c\overline{c}$ momentum. 
The selection criterion is that the angle 
of $c\overline{c}$ is smaller than $110^{\circ}$. 
Solid line: no CS states; 
Dashed and dotted line: 100\% and 30\% CS states, respectively. 
(d) Rapidity distribution of charged particles for c-events. 
The rapidity axis is chosen along the added 
$c\overline{c}$ momenta. The selection criterion is that the angle 
of $c\overline{c}$ is smaller than $110^{\circ}$. 
Solid line: no CS states; Dashed line: 30\% CS states. }
\label{fig:3}
%\vspace*{.4in}
\end{figure}

\begin{figure}
%\vspace*{1in}
 \setlength{\epsfxsize}{6in}
  \centerline{\epsffile{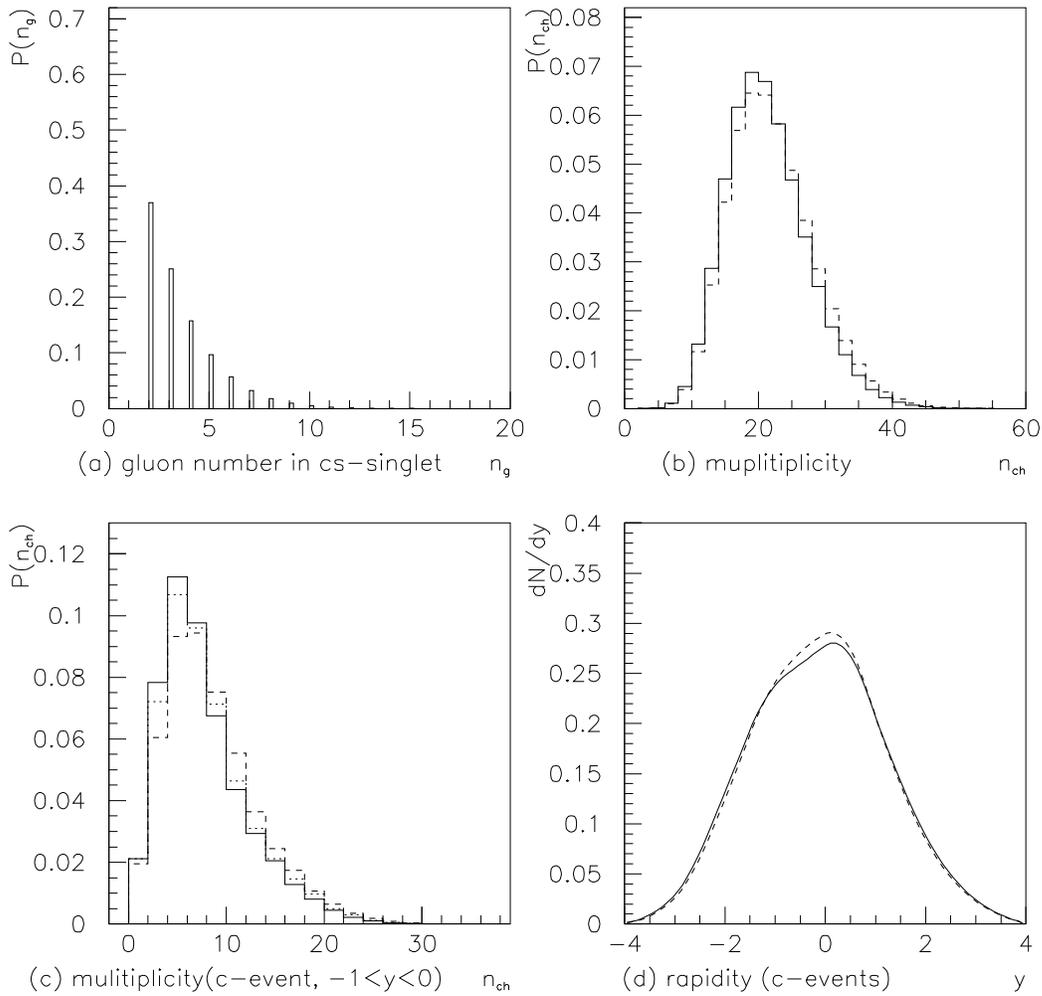}}
%\vspace*{.5in}
\caption{
The same as Fig.\ref{fig:3} except that the constant 
weight is used in selecting CS states.}
\label{fig:4}
%\vspace*{.4in}
\end{figure}

Now we begin to discuss our Monte Carlo results which are shown 
in Fig.\ref{fig:1}-\ref{fig:4}. Fig.\ref{fig:1} and \ref{fig:2} 
show how CS states influence the global properties of events, 
while Fig.\ref{fig:3}(a-d) and \ref{fig:4}(a-d) shows other properties in the 
case of the T-weight and the constant weight 
respectively: the distribution of the 
gluon number in the gluonic sub-singlet of a CS state, the multiplicity 
distribution in the whole and the restricted phase space, and the rapidity 
distribution for c-events in the central region. 
The constant weight is defined as that each 
possible CS state is chosen with 
the same probability, and the T-weight is discussed 
earlier in this section. We show three 
groups of results: no CS states, 100\% and 30\% CS states. 
These results show that for the situation of the constant weight, 
there is observable though small deviation between 
the result with CS states and that without CS states, while for 
the situation of the T-weight, the deviation is even smaller. 
One may easily accredit the small deviation in event properties to 
the $1/N_c^2$ effect of the CS state. This is not truely the case. 
Here two questions should be distinguished: 
(a) How large is the pobability for CS states to appear?
and (b) How large is the deviation between the hadronic events 
with and without CS states included? 
They are two distinct concepts by nature. 
We recall that there are many non-orthogonal leading 
CS states corresponding to an ordered trace state. Though a 
single leading CS state contribute only $1/N_c^2$, the total contribution 
may actually be large for many CS states. 
In fact we cannot determine the total probability at the 
perturbation level as shown earlier in the former section.
Hence we include a parameter R to denote the total probability of CS states 
which we will give results in two cases R=30% and 100%. 
Note that all the above statements are refer to question (a). 
But when we talk about the small deviation between 
events with and without CS states,  we are addressing question (b). 
One can understand this deviation as the 
result of the soft gluon dominance in the 
cascade and the infared property of global observables, considering that 
the global shapes are mainly determined by the cascade instead of the
hadronization and that the effect caused by 
isolating a group of gluons (mostly soft continuous gluons in 
the color sequence) should not be large. 
We know earlier in this section and later in the 
following paragraph that T-weight favors 
the CS states with softer and less gluons in their gluonic subsinglets which 
bring little change to global and local observables, while 
the constant weight gives each CS state equal probability which 
leads to larger contribution from those CS states with harder and 
more gluons in their gluonic subsinglets whose observables 
may differ significantly from those conventional 
events without CS states. So on average 
the constant weight gives larger difference from conventional events 
than the T-weight. 

It is meaningful to investigate local observables especially in 
carefully chosen phase space regions and types of events, where 
the difference between CS events and Non-CS ones may 
be more clearly seen. Fig.\ref{fig:3}a shows the probability distribution of 
gluon number in the gluon type sub-singlet of a CS 
state. The T-weight sampling is used to select 
a CS state for a given parton configuration. 
The distribution shows a perfect exponential decreasing feature: 
$P(n_g)\sim exp(-0.9n_g)$. The mean value of the gluon number 
$<n_g>\sim 2.9$. Fig.\ref{fig:4}a is the $n_g$ distribution with 
the constant weight. Its decreasing behavior is much slower 
than that with the T-weight. It has the mean value $<n_g>\sim 3.5$. 
We know that the T-weight favors the CS state with soft gluons in its gluonic subsinglet. 
More gluons are in its gluonic subsinglet, 
more probably it gives a smaller T-weight. 
Hence the T-weight favors the smaller number of gluons 
in the gluonic subsinglet of a CS state. 
Therefore the $<n_g>$ distribution with the T-weight 
is steeper than that with the constant weight. 
Fig.\ref{fig:3}b is the multiplicity distribution for charged particles using 
the T-weight. The solid line is the result without CS states, 
and the dashed line is that CS states are produced 
with the total fraction of 30\%. We see that 
there is little difference after introducing CS states. 
It is not surprising because we know that in most cases 
the gluonic sub-singlet of a CS state is composed of soft gluons, which 
brings little change to the multiplicity of the original string connection. 
Fig.\ref{fig:4}b is the multiplicity distribution for charged particles in 
the constant weight. An apparent deviation from that 
without CS states can be seen. The broader distribution 
gives a larger mean multiplicity than in the T-weight case. 
This phenomenon is consistent with our above analysis that 
compared to the T-weight case, the constant 
weight enhances the probability of CS states 
with harder gluonic subsinglets which give rise to larger multiplicity. 
Fig.\ref{fig:3}c is the multiplicity distribution of charged particles in 
the T-weight in the rapidity range $-1<y<0$ for the c-events. 
The rapidity axis is chosen along the 
added $c\overline{c}$ momentum. The selection criterion is that the angle 
of $c\overline{c}$ is smaller than $110^{\circ}$. The solid line 
is the result without CS states, and the dashed and dotted line are 
that CS states are produced with total fractions of 100\% and 30\%, 
respectively. The reason for choosing the c-event is that 
for these events it is easy to tag the primary quarks. Normally there should 
be a rapidity gap caused by CS states between 
the $c\overline{c}$ cone and the cone in the opposite direction. 
That is to say, in the rapidity region around $y\sim 0$, there are 
fewer particles than the case without CS states. 
This effect can be seen in the figure. 
The probability to have fewer than 2 charged particles in the range 
$-1<y<0$ in the case with CS states produced is obviously larger 
than that without them. This is just what we expect 
a sign (though a weak one) of a depleting area or a rapidity gap. 
Fig.\ref{fig:4}c is the result with the constant weight. 
No sign (even a weak one) of a depleting area is found. 
This is possibly due to the broader multiplicity distribution brought by 
the constant weighting (see also Fig.\ref{fig:4}b). 
Fig.\ref{fig:3}d is the rapidity distribution of 
charged particles for the c-event in the T-weight. 
The rapidity axis is chosen along the added 
$c\overline{c}$ momenta. The selection criterion is that the angle 
of $c\overline{c}$ is smaller than $110^{\circ}$. 
The solid line is the result without CS states, and the dashed line is 
that CS states are produced with total fraction of 30\%. 
The difference of the largest and the most interest is in the region $-2<y<0$, 
where the distribution with CS states is a 
little lower than that without them. 
This is a weak indication for the particle defeciency 
in the central region. 
This phenomenon is absent for Fig.\ref{fig:4}d where 
the constant weight is used. On the contrary, 
the central region of the distribution with CS states is clearly higher 
than that without them, which implies a more 
populated central area exists for the case of CS states. 

In the literature, there are a number of 
reconnection models which can describe 
the production of CS states in a variety of reactions 
\cite{friberg,torb,leif}. Ref.\cite{leif} gives a reconnection 
model(which we call RMDCM for short) 
which is incorporated into the dipole cascade Monte Carlo 
ARIADNE. In some sense, 
its strategy of color reconnection is similar to ours, 
but it is more complicated and more phenomenologically based. 
Our model provide a significantly simpler recipe for 
reconnection. The results for $e^+e^-$ annihilation 
produced by our model and RMDCM are similar. 

Noticing that the CS states with soft-gluon sub-singlet have large 
T-weights and hence have large production rates, the results of our model 
have some similarities with those from the 
Soft Color Interaction (SCI) model\cite{edin}. 
We will in the future study implications from our model for 
deep inelastic scattering.

\section{Summary}

We use the method of color effective Hamiltonian $H_c$ to study 
properties of color separate states in 
$e^+e^-\rightarrow q\overline{q}+ng$. 
The full physical color state is $\left| H_c \right\rangle$.
There are $n!$ trace terms in $\left| H_c \right\rangle$, each 
corresponds to an order of $n$ gluons. 
We call each trace term a trace state. 
Up to $O(1/N_c^2)$, there are $m=\frac 12n(n-1)$ CSGSs 
with the leading contribution to a trace state. 
We obtain the following rules for a leading CSGS:
Up to the leading order in $1/N_c$, 
(a) The state has only two sub-singlets. Each sub-singlet 
keeps its parton order as that in the trace state at most. 
(b) Its projection square on the trace state is $1/N_c^2$.
(c) The inner product of any two different 
leading states is $1/N_c^2$, while that of a leading state with a 
non-leading one is $1/N_c^3$ or higher. 

Due to the non-orthogonality of CSGSs, 
it is necessary to orthogonalize 
them to give the total probability. Actually 
the general orthogonalization proves 
to be extremely difficult.
But it is simple and heuristic 
in some circumstances, e.g. 
the physical state is fixed at one trace state. 
In this case, we can estimate the probability 
of the CSGS through orthogonalization under 
some further approximations:
(a) We take into account leading CSGSs and 
the singlet chain state with the same gluon order as 
that of the trace state.
(b) We choose a simple form for the transformation matrix.
The orthogonalization procedure has two steps. 
The first step is to find the orthogonal 
states $\{|s_i'\rangle ,i=1,\cdots ,m\}$ from 
$m$ leading CSGSs $\{|s_i\rangle ,i=1,\cdots ,m\}$. 
The second step is to introduce the singlet chain state 
$\{|f\rangle \}$ which is approximately the trace 
state $\{|h\rangle \}$ into the list, and to obtain 
the final set of orthogonal states: 
$\{|f'\rangle ,\{|s_i''\rangle ,i=1,\cdots ,m\}\}$. 
In the large $N_c$ limit, the probabilities 
of $|f^{\prime }\rangle$ and 
one $|s_i^{\prime \prime}\rangle $ in the corresponding 
normalized trace state $| h\rangle $ are given by:
$P_{f'}\simeq \frac{1}{1+m/N_c^2}$ and 
$P_{s''}\simeq (1-\frac{1}{1+m/N_c^2})/m\simeq 1/N_c^2$. 
The total probability for all $|s''\rangle $ states 
are $mP_{s''}\simeq 1-\frac{1}{1+m/N_c^2}\simeq m/N_c^2$. 
We see that $P_{f'}$ decreases while $mP_{s''}$ increases 
monotonously with the growing gluon number(or $m$), 
and $P_{f'}+mP_{s''}$ conserves the total probability 1.
These results are for large $N_c$. When $N_c$ is finite, 
we always span the pole at $m=N_c^2/2$ for $P_{s_i''}$ 
which is unphysical. Hence the present result is 
invalid for finite $N_c$. In this case, We have to 
consider all non leading terms in $N_c$ 
which is too complicated to have an explicit result. 

The rules for the leading CSGS are crucial for us to build a 
practical model to describe the production of color 
separate states in event generators. Here are the 
outline of the model:
\begin{enumerate}
\item  We use a parameter $R$ to describe the total fraction of CS 
states. We use JETSET to produce a parton 
configuration with a specific parton connection. We let it 
correspond to a trace state.

\item We only consider CSGSs with the leading contribution 
to the trace state. 
According to the cascade history recorded by JETSET 
for the current configuration, we can further exclude CS states 
whose gluon-type sub-singlets are emitted from a single 
octet gluon. 

\item We consider two kinds of weights in selecting individual 
CS state for an event. One is the constant weight. 
This is a natural choice if we only consider the color part and 
neglect the momentum or phase space sector.
The other is the T(or $\lambda$) measure weight.

\item The fragmentation of the selected CS state is implemented by 
the Lund string model. Note that the gluon 
sub-singlet forms a closed string whose 
fragmentation should be managed with care. 
\end{enumerate}

If we use $T$ measure weight, we find that most of CS states 
produced in the process have the following configurations: 
(a) The gluon-type sub-singlet of 
a CS state is made up of all gluons in several continuous branches. 
If part of a branch joins the sub-singlet,  
other part of this branch should also join it. 
Any separation of one part from the rest of the same branch would 
result in the suppression of the production weight. 
(b) CS states with their gluon-type sub-singlets formed by adjacent 
twigs of the branching tree have large production rate.
From these points, we see that, CS states whose gluon-type 
sub-singlets consist of soft adjacent gluons 
have relatively larger T-weights than other 
types of CS states and therefore have larger production rates. 

The comparisons of global and local event properties between events with 
and without CS states in the T- and the constant weight 
are given. Generally the difference of results between two cases, 
those with and without CS states, is found to be small, though we find 
the deviation in the constant weight is a little larger than that in the T-weight, 
and in the central phase space area a weak sign of the 
rapidity gap is shown in the T-weight case. 
The main reason lies in the fact 
that soft gluons dominate in the cascade. Considering the infared 
safe properties for all observables, the effect caused by 
isolating a group of gluons (mostly soft continuous gluons in 
the color sequence) should not be large. 
T-weight favors CS states with soft gluonic sub-singlets which 
should not bring significant change to global and local observables, while 
the constant weight gives each CS state equal probability which 
enhances contributions from those CS states with harder 
gluons in their sub-singlets whose observables may differ significantly from conventional 
states without separate subsinglets. So on average the constant 
weight gives rise to larger deviation from conventional events than the T-weight. 
Conclusively we find that in the high energy electron-positron 
annihilation, global and local properties are mainly determined by the 
momentum configuration of final partons, and are almost insensitive to the way of 
color connection. 

{\bf Acknowledgement}
One of authors Q. Wang is grateful to Prof. Andersson for his kind 
invitation for visiting the Department of Theoretical Physics, 
Lund University where this work was done and where 
he learnt from Prof. Andersson a lot about the T-measure 
and much other knowledge about 
the dipole cascade model. Also, many thanks from Q. Wang to 
Prof. Sj\"ostrand for his kind help with JETSET and 
computer programming, Leif L\"onnblad, Patrik Ed\'en for their helps on 
ARIADNE, and Christer Friberg for helping him solve a lot of 
computing problems. This work is supported in part by 
National Natural Science Foundation of China. 

\newpage

\end{document}